# Video Streaming Over QUIC: A Comprehensive Study

JASHANJOT SINGH SIDHU, Concordia University, Canada
ABDELHAK BENTALEB, Concordia University, Canada

The QUIC transport protocol represents a significant evolution in web transport technologies, offering improved performance and reduced latency compared to traditional protocols like TCP. Given the growing number of QUIC implementations, understanding their performance, particularly in video streaming contexts, is essential. This paper presents a comprehensive analysis of various QUIC implementations, focusing on their transport-layer congestion control (CC) performance and its impact on HTTP Adaptive Streaming (HAS) in single-server, multi-client environments. Through extensive trace-driven experiments, we explore how different QUIC CCs impact adaptive bitrate (ABR) algorithms in two video streaming scenarios: video-on-demand (VoD) and low-latency live streaming (LLL). Our study aims to shed light on the impact of QUIC CC implementations, queuing strategies, and cooperative versus competitive dynamics of QUIC streams on user QoE under diverse network conditions. Our results demonstrate that identical CC algorithms across different QUIC implementations can lead to significant performance variations, directly impacting the QoE of video streaming sessions. These findings offer valuable insights into the effectiveness of various QUIC implementations and their implications for optimizing QoE, underscoring the need for intelligent cross-layer designs that integrate QUIC CC and ABR schemes to enhance overall streaming performance.

CCS Concepts: • **Networks** → **Transport protocols**; • **Information systems** → **Multimedia streaming**.

Additional Key Words and Phrases: Video Streaming, QUIC, HTTP Adaptive Streaming, Congestion Control.

## 1 Introduction

According to the latest video streaming report by Ericsson [20], video streaming is projected to account for 74% of mobile data traffic by the end of 2024, underscoring the dominance of video streaming traffic across the Internet. Major Over-The-Top (OTT) platforms like Amazon Prime Video, Netflix, Meta, and YouTube significantly contribute to this massive data consumption, relying on HTTP Adaptive Streaming (HAS) technologies [10] such as MPEG Dynamic Adaptive Streaming over HTTP (DASH) and Apple HTTP Live Streaming (HLS) [56] to deliver their contents and adapt to variable network conditions thanks to Adaptive BitRate (ABR). Transmission Control Protocol (TCP) [15] was the standard transport protocol for delivering HAS-based traffic. In the last three years, the advent of QUIC transport protocol, originally proposed by Google [41] represents a significant advancement in network communication and media delivery [23]. Platforms such as YouTube and Facebook have increasingly adopted HTTP/3, which is built on top of QUIC, to deliver video traffic. This protocol introduces several key features, including multiplexing, 0-RTT,

Authors' Contact Information: Jashanjot Singh Sidhu, Concordia University, Montreal, Canada; Abdelhak Bentaleb, Concordia University, Montreal, Canada.

connection establishment, prioritization, and seamless connection migration. These capabilities make QUIC a robust alternative to TCP, especially for HAS-based traffic. By operating over UDP and serving as the transport layer for HTTP/3, QUIC improves video streaming efficiency and security, leading to enhanced Quality of Experience (QoE) compared to traditional TCP-based methods [36, 41, 61, 68]. Since HTTP/3 is fully built on top of QUIC, HAS traffic today effectively depends on both layers: the HTTP/3 layer for content delivery semantics, and the QUIC transport layer for congestion control, multiplexing, and reliability. In this work, we primarily study QUIC's transport-layer mechanisms, particularly congestion control, while evaluating their effect on HAS-based video delivery.

The Internet Engineering Task Force (IETF) played a crucial role in the refinement and standardization of QUIC, a protocol initially introduced by Google in 2012. Over nearly a decade of development and iterations, QUIC matured and was formally adopted as the foundation of the HTTP/3 protocol by 2021 [39]. The standardization of QUIC by the IETF led major technology companies, including Meta, Microsoft, and Tencent, as well as other industry leaders like Google, Cloudflare, and Akamai, to develop their implementations of the protocol. The broad adoption and customization of QUIC by these industries underscores its growing significance in modern Internet infrastructure. Although QUIC has been standardized, a variety of independent implementations have been developed. Each implementation emphasizes different design goals—such as performance, security, or CDN integration—while remaining compliant with the QUIC standard to ensure interoperability. For example, while Google's QUIC implementation is written in C++ and focuses on performance and integration with Google's services, Cloudflare's implementation is in Rust and emphasizes security and CDN-specific optimizations. Similarly, Microsoft and Akamai have their own C++ implementations tailored to their respective use cases, and Meta also has a version adapted for its video streaming needs. These diverse implementations not only reflect the flexibility of the QUIC protocol but also highlight the absence of a one-size-fits-all solution, as each implementation incorporates unique features, congestion control algorithms, and performance enhancements based on specific requirements.

Typically, a HAS-based server-side congestion control (CC) algorithm [52] manages the sending rate according to the network capacity. Meanwhile, on the HAS-based client side, an ABR [10] algorithm determines the best representation (bitrate and resolution pair) for future segments based on factors such as available bandwidth and playback buffer levels. Effective CCs are crucial for ensuring a smooth streaming experience and high QoE by optimizing data transmission rates (higher bitrate) and minimizing interruptions and delays (lower rebuffering events). However, the diversity in QUIC implementations [22], such as QUICHE, MVFST, QUINN and AIOQUIC, each with different CCs and varying implementations, significantly impacts the QoE (Section 4). For instance, QUIC implementations like QUICHE use HyStart [24] to mitigate the negative effects of the slow start in Cubic. However, even when two implementations, such as TQUIC and QUICHE, employ the same HyStart mechanism, their delivery rate mechanisms vary due to differences in design choices, programming languages, and optimization priorities. Additionally, the design of these QUIC implementations to support multiple clients simultaneously varies. For example, QUINN uses asynchronous API calls, TQUIC employs multi-path QUIC (MPQUIC) [44] or multiple threads, while QUICHE lacks such support altogether. This absence of a unified implementation adaptable to specific networking and application scenarios is driving the development of Media over QUIC (MoQ) stack [29], which is in its initial development stage.

The variations in QUIC implementations can significantly impact QoE for video streaming, particularly as QUIC servers are scaled for diverse real-world scenarios. To our knowledge, this is the first comprehensive analysis of how implementation differences in QUIC affect CC algorithms and, consequently, ABR and QoE performance. Through a series of detailed trace-driven emulations using the Vegvisir framework [26], the paper evaluates single-server multi-client (SS-MC) setups with varied real-world network traces and video contents for both VoD and LLL modes. The insights gained are essential for optimizing video streaming performance and improving QoE, offering actionable recommendations for selecting and configuring QUIC implementations to ensure efficient, high-quality content delivery. This paper offers a holistic and comprehensive evaluation of HAS over QUIC, encompassing various settings such as network conditions, video contents, CCs and ABR algorithms. It mainly presents four key contributions:

(1) We rigorously evaluate the performance of various QUIC CCs across seven implementations, including AIOQUIC, MVFST, LSQUIC, PICOQUIC, QUINN, TQUIC, and XQUIC.
(2) We perform an in-depth analysis of the effect of QUIC+CC on HAS in various network settings, revealing the intricate relationship between CCs and QoE.
(3) We provide a detailed examination of how Active Queue Management (AQM) at forwarding devices (routers) influences QUIC+CC performance, alongside the impact of fluctuations in 5G/4G networks. Additionally, we investigate the role of Explicit Congestion Notification (ECN) and examine the coexistence of QUIC flows with TCP traffic. This analysis offers valuable insights for a plethora of real-world use cases, highlighting how various network management strategies and conditions can shape the performance of QUIC in dynamic environments.
(4) Our experimental results demonstrate that different QUIC+CCs lead to significantly varied QoE outcomes for video streaming sessions. Each QUIC implementation exhibits distinct strengths and weaknesses, making them more or less optimal for specific scenarios, thus providing crucial guidance for implementation selection and tuning.

## 2 Related Work

The QUIC interop runner [62] performs a range of tests on multiple QUIC implementations, including scenarios such as blackhole, handshake corruption, cross-traffic, long RTT, and IPv6. While these tests provide valuable insights into QUIC+CC and goodput results, they do not address video streaming scenarios or evaluate QUIC+CC performance across different network conditions. Additionally, they do not assess the impact of QUIC+CC on HAS streaming. Furthermore, the interop runner primarily employs the default CCs within each QUIC implementation and does not compare various CCs or their implementation differences. Zhang *et al.* [69] assessed the performance of QUIC on high-speed networks compared to TCP+TLS+HTTP/2. Kempf *et al.* [38] extended the QUIC Interop Runner to benchmark QUIC performance on high-speed 10G links and highlighted issues with default buffer sizes, inefficient packet I/O, and sub-optimal use of hardware acceleration. Kunze *et al.* [40] monitored the sending behavior of QUIC flows and assessed their responsiveness to congestion signals such as packet loss and ECN markings. Dey *et al.* [19] provided a critical comparative study of TCP/IP and the emerging QUIC protocol, focused on connection establishment, speed, reliability, security and highlighted the advantages and vulnerabilities of QUIC. Bhat et al. [11] provided one of the first systematic evaluations of DASH over QUIC, highlighting both benefits and limitations compared to TCP-based streaming. Palmer et al. [55] explored optimizations for video delivery over QUIC, emphasizing performance and interoperability challenges. Mazhar and Shafiq [47] developed a real-time QoE monitoring framework for both HTTPS and QUIC traffic, emphasizing the importance of end-user experience. Mondal and Chakraborty [49] studied whether QUIC aligns well with modern ABR techniques, showing that transport–application interactions play a critical role. Most recently, Bottisio [13] provided an in-depth doctoral study on media streaming performance over HTTP/3. While this work emphasizes performance analysis within the standardized HTTP/3 stack, our focus is on comparing diverse QUIC transport-layer implementations and their congestion control mechanisms, thereby complementing Bottisio's HTTP/3-centric investigation.

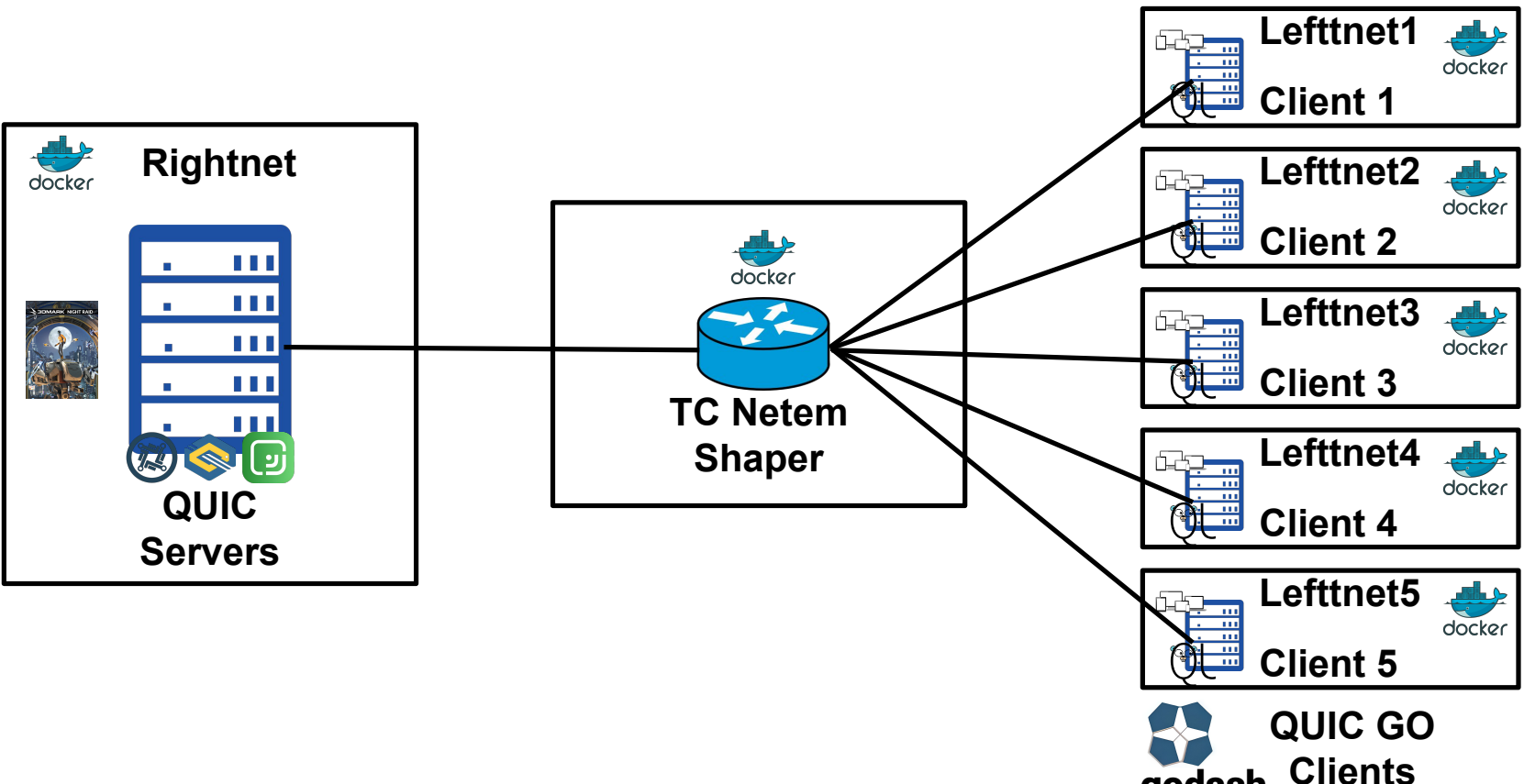

Fig. 1. Extended Vegvisir Emulation Setup.

Previous studies primarily examined general QUIC performance, high-speed networks, or specific aspects such as congestion responsiveness. However, they relied on the default CCs within each QUIC implementation and did not evaluate the scalability of QUIC+CC in video streaming scenarios. In contrast, our work systematically evaluates multiple QUIC+CC implementations and their impact on ABR algorithms and QoE under diverse network conditions. We provide a comparative analysis of how different CCs influence QoE in HAS, thereby addressing gaps left by prior research.

## 3 Methodology

We extended the Vegvisir evaluation framework [26] to support a single server setup with one server connected to five clients (denoted as SS-MC) through a single router (referred to as a shaper) running tc Netem [25], creating a one-to-many mapping where all clients simultaneously request contents, as illustrated in Figure 1. We emulated typical HAS video streaming sessions for both video-on-demand (VoD) and low-latency-live (LLL) streaming modes using a 4K video across various network conditions. The QUIC-GO client utilized a modified goDASH-based player from [27], adapted to operate directly over QUIC as the transport layer, in headless mode, which simplified metric extraction compared to using a web-based dash.js player [21]. Network emulation was performed using tc-Netem with various traces, as depicted in Figure 2. The bottleneck link at the shaper was configured with a total capacity equal to five times the bandwidth of a single client's trace, accurately simulating shared resource contention among the five clients. For detailed configurations used in all scenarios, refer to Table 1, where D=description (*col*. 3) denotes whether all the clients are operating under same or different network traces with same or different ABR. The segment durations and maximum buffer sizes in Table 1 were chosen based on typical values reported in prior work [35, 63, 64]. For VoD, a 4s segment with a 60s buffer reflects common deployment settings where larger buffers smooth throughput fluctuations and allow pre-fetching of higher-quality segments. For LLL, a 2s segment with a 6s buffer matches low-latency live streaming constraints, where shorter buffers reduce latency and enable faster ABR decisions. While we report only these representative values for clarity, preliminary tests with alternative segment lengths and buffer sizes showed similar trends in ABR behavior and QoE metrics, confirming that these settings effectively capture the performance differences between VoD and LLL. Note that for the shaper, the default AQM [2] strategy *Packet First-In, First-Out* (PFIFO) [48] was used. The server communicates with the traffic shaper over a shared subnet, using a Docker bridge network defined by Docker Compose. This bridge facilitates seamless and isolated communication between components. The traffic shaper manages data flow using Token Bucket Filters (TBF) to simulate various network conditions, such as delay, loss, and bandwidth constraints. Traffic from the shaper is distributed to multiple clients, each running in a separate Docker container with dedicated connectivity. A corresponding Docker bridge on the client side ensures proper routing. All containers — server, shaper, and clients — operate within the same virtual subnet (`193.167.0.0/16`), enabling consistent and reproducible data flow across the testbed. Additional implementation details can be found in Appendix D.

Our setup comprised a physical machine running Ubuntu 22.04.4 LTS, NVIDIA GeForce RTX 3060 24-core CPU, and 64 GB memory. During the testing phase, the client, shaper, and server were each encapsulated within Docker containers, forming a cohesive end-to-end video streaming system. These containers were interconnected through six Docker-compose networks for SS-MC, orchestrated to pass through the shaper, effectively managing and regulating the flow of traffic between them, as shown in Figure 1.

Table 1. Details of experimental scenarios.

| Mode | Scen. | Description | Seg. Dur. (s) / Max. Buff. (s) |
|---|---|---|---|
| **VoD** | A | Same trace, Same ABR | 4 / 60 |
| | B | Same trace, Different ABR | |
| | C | Different trace, Same ABR | |
| **LLL** | D | Same trace, Same ABR | 2 / 6 |
| | E | Same trace, Different ABR | |
| | F | Different trace, Same ABR | |

*3.0.1 Video Sample and Parameters. Moment of Intensity* (MoI) video [1] was used during the evaluation with different segment duration and maximum buffer capacity. MOI is a 4K video with 3840x2160 maximum resolution and was encoded using a bitrate ladder ranging from 0.5 Mbps to 40 Mbps. We tested the efficiency of different CC across QUIC implementations using two essential streaming modes: video-on-demand (VoD) and low-latency-live (LLL), with multiple scenarios for each mode. To ensure accuracy, the video streaming session ran for approximately 100 seconds for each video. While we experimented with longer durations, 100 seconds was deemed sufficient to capture the essential dynamics between the QUIC+CC implementations. This duration allowed us to observe key performance metrics such as throughput, delay, and the impact of different CCs, without unnecessarily prolonging the tests. Additionally, given the large volume of data generated (~70 GB) for a single scenario due to 4K content and qlogs [46]—this time frame provided a balance between capturing meaningful results and managing data storage constraints.

*3.0.2 Network Traces.* We used three network traces, 5G Netflix [58], LTE Belgium [67] and Cascade network trace with sampling intervals of 1s for the first two traces, and ~30s for the last. Note that the 5G traces are sub-6GHz and not mmWave. All the traces are illustrated in Figure 2.

*3.0.3 CC Algorithms and Metrics.* Table 2 presents a comprehensive summary of existing QUIC implementations, detailing their supported CC algorithms and known limitations. Notably, only a subset of these implementations offer stable support for both multi-client scenarios and CC algorithms without critical bugs. For our SS-MC experimental setup, we integrated AIOQUIC, MVFST, PICOQUIC, LSQUIC, QUINN, TQUIC, and XQUIC as QUIC servers, each with their respective CC support [22]. We deliberately included XQUIC and LSQUIC, despite their lack of multi-client support, to underscore the significant impact that this limitation has on the QoE for clients in HAS scenarios. As shown in Table 2, AIOQUIC supports Cubic and Reno; MVFST supports BBR, BBR2, Cubic, Reno, Copa and Copa2; PICOQUIC supports BBR, BBR3, Cubic, Reno, DCubic, FastCC and Prague; QUINN supports BBR, Cubic and Reno; TQUIC offers BBR3, BBR, Copa, and Cubic; and XQUIC offers BBR, BBR2, Cubic, Reno and Copa. To assess the performance of these CCs, we focused on two key metrics: average throughput and delay. These values were extracted from their respective qlog files. For implementations like AIOQUIC and QUINN, which do not natively support qlogs, we modified their implementations to log these metrics directly from their respective CC classes. The specs for each of the used CC can be found at [1]. Notably, DCubic in PICOQUIC is a delay-based variant of Cubic that exits slow start upon detecting prolonged delays or significant packet losses. In addition, Table 3 provides an overview of the QUIC stack features across these implementations (e.g., concurrency model, pacing, multipath, ECN, and TLS backend). We relied on the default feature set for each implementation, with multipath explicitly disabled where present, to ensure comparability. This overview clarifies which stack-level differences could influence streaming performance, in line with the QUIC interop matrix, and directly addresses potential sources of variation beyond congestion control.

[1] https://youtu.be/JLTmjZmqiNc

*3.0.4 ABR Algorithms and Metrics.* We used three heuristic-based and two learning-based ABR algorithms. Among heuristic-based algorithms, we have one buffer-based, namely BBA2-XLDouble (denoted as BBA2-C), and two throughput-based, namely CON (Conventional) and EXP (Exponential), all from [27]. BBA2-C incorporates an additional stall prevention mechanism to proactively reduce playback stalls. CON uses the average of the last two throughput estimation samples to perform ABR decisions whereas EXP uses the exponential moving average of the most recent 10 throughput estimation samples to perform ABR decisions. For the learning-based ABR, Pensieve [45] and Merina [37] were used. Originally, Pensieve and Merina are trained to choose among 6 bitrate representations but since our video samples consist of 10 bitrate levels, we retrained them for 10 bitrate levels and with more epochs. Additionally, we also implemented the stall logic of BBA2-C in Pensieve (denoted by Pensieve$^{+}$) and Merina (denoted by Merina$^{+}$).

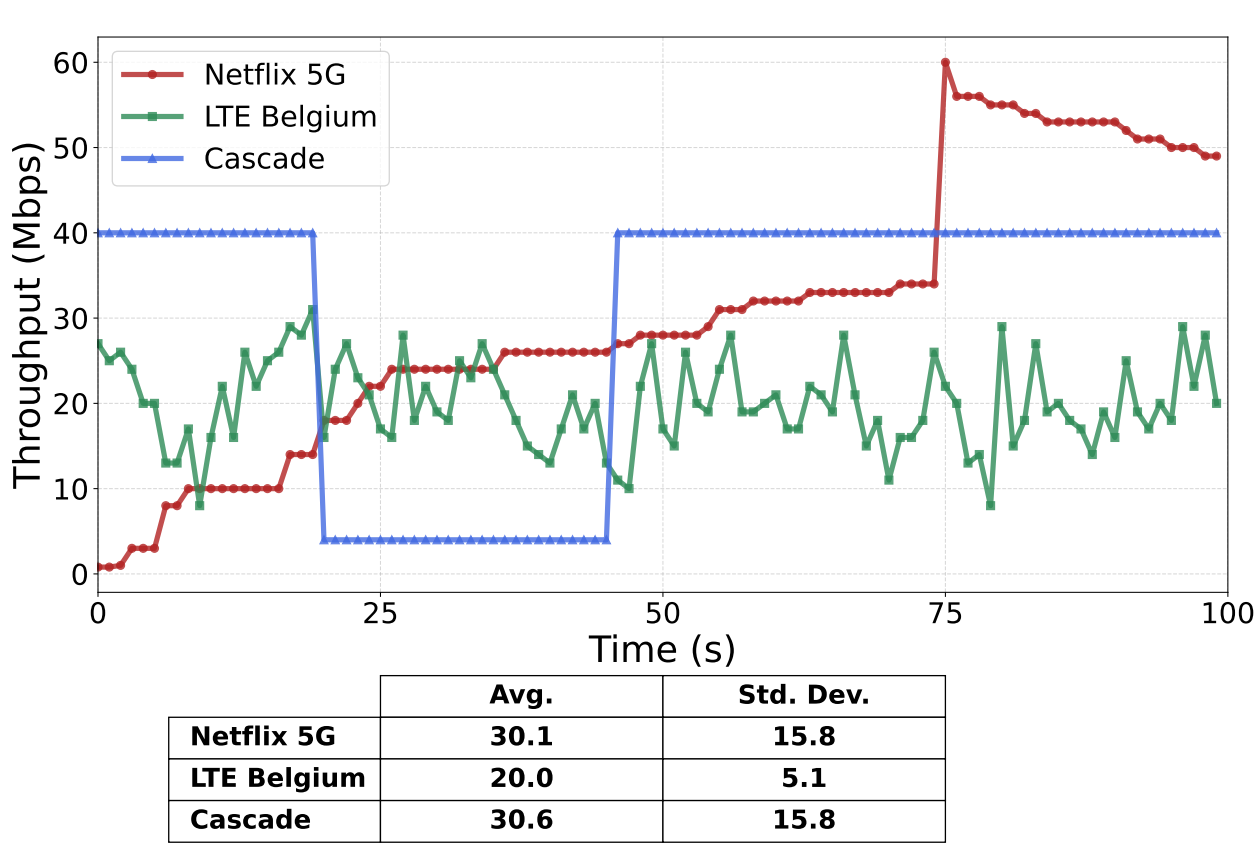

| | Avg. | Std. Dev. |
|---|---|---|
| **Netflix 5G** | **30.1** | **15.8** |
| **LTE Belgium** | **20.0** | **5.1** |
| **Cascade** | **30.6** | **15.8** |

Fig. 2. Network Traces.

To have comprehensive comparison amongst combination of all algorithms (CC+ABR) for each QUIC implementation, we used VMAF [43] as a perceptual quality metric along with rebuffering duration ratio (RD) as these key metrics directly influence the QoE of a video streaming session. Please note that we opted to exclude the widely adopted QoE$^{itu}$ [33] model as the ITU-P.1203 implementation is limited to 1080p content and the extended ITU-P.1204 implementation is still experimental, lacks frame-level parsing, and is limited to short 10s videos [14, 65].

## 4 Performance Evaluation

In this section, we present the performance results of various QUIC implementations for scenarios **A** and **D** (outlined in Table 1). The evaluation covers both VoD and LLL modes under diverse network conditions. Notably, we did not observe major rebuffering events in either VoD or LLL modes. Most QUIC+CC combinations fell into one of two extremes: either they achieved zero or minimal rebuffering, or they experienced excessive rebuffering that prevented all clients from being served simultaneously. This is evident from the empty VMAF bars corresponding to these combinations, which indicate their inability to deliver consistent quality to all clients. Hence, we divert our focus toward VMAF as an evaluator for video streaming QoE. Note that the results from scenarios **B**, **C**, **E**, and **F** align closely with those from scenarios **A** and **D** and thus, the results for these scenarios are included in Appendix B.

Table 2. Summary of QUIC implementations.

| Impl. | CC | Multi-Client | ECN [60] | Issues |
|---|---|---|---|---|
| Aioquic | Cubic, Reno | Threads | Y | Long RTT times in multi-client setups |
| Go-x-net | - | Threads | N | No CC support |
| Haproxy | Cubic, Reno, BBR | Threads | Y | No client support |
| Kwik | Reno | Threads | N | Frequent data transmission can lead to connection interruption [57] |
| Lsquic | Cubic, BBR, Adaptive | NA | Y | Relies on only a single thread [4] |
| Mvfst | Cubic, Reno, BBR, BBR2, Copa, Copa2 | Asyncio [31] | N | Ignores ACK and does not send anti-deadlock packets [32] |
| Msquic | Cubic, BBR | Threads | N | Multi Connection issues [53] |
| Neqo | Cubic, Reno | Threads | N | QUIC Stream Resets with Partial Delivery [50] |
| Ngctcp2 | Cubic, Reno, BBR | Threads | Y | No support for multi-path QUIC |
| Nginx | - | Threads | N | No CC support; No client support |
| Picoquic | Cubic, Reno, BBR, BBR2, DCubic, FastCC | Threads | Y | ACK Trimming is too aggressive [28] |
| Quic-go | Cubic | NA | N | SendDatagram blocks all calls under extreme congestion [17] |
| Quiche | Cubic, Reno, BBR, BBR2 | NA | N | No support for multiple connections, sends data in a for loop per stream [30] |
| Quinn | Cubic, Reno, BBR | Asyncio [18] | N | Loses packet states in low throughput conditions [12] |
| S2n-quic | Cubic, BBR | Threads | N | Multi-Client support is platform dependent |
| Tquic | Cubic, BBR, BBR3, Copa | Threads | Y | Missing pacing support [66] |
| Xquic | Cubic, Reno, BBR, BBR2, Copa | NA | N | No asynchronous SSL management, re-writes HTT3 context with new request [5–7] |

Table 3. Overview of QUIC stack features considered in our evaluation.

| Impl. | Language | Concurrency Model | Pacing | Multipath | ECN | Qlog/Logging | TLS/SSL |
|---|---|---|---|---|---|---|---|
| Aioquic | Python | Threads | ✗ | ✗ | ✓ | Custom | OpenSSL |
| Lsquic | C | Single-threaded | ✓ | ✗ | ✓ | ✓ | BoringSSL |
| Mvfst | C++ | Async I/O (folly) | ✓ | ✗ | ✗ | ✓ | Fizz (Meta TLS) |
| Picoquic | C | Threads | ✓ | ✓ | ✓ | ✓ | OpenSSL |
| Quinn | Rust | Async I/O (tokio) | ✓ | ✗ | ✗ | Custom | rustls |
| Tquic | Rust | Multi-threaded | ✗(ACK clocking) | ✓ | ✓ | ✓ | rustls |
| Xquic | C | Single-threaded | Partial | ✗ | ✗ | ✓ | BoringSSL |

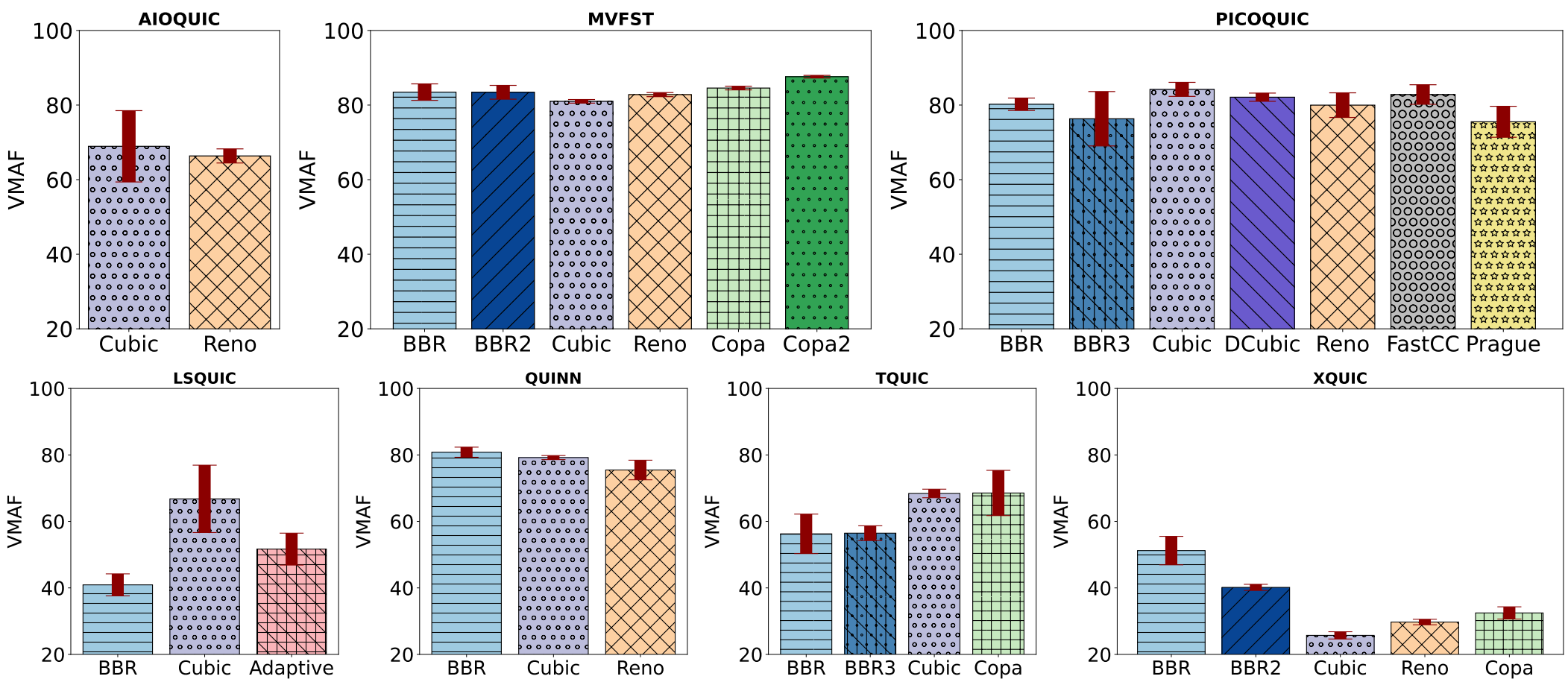

Fig. 3. Avg. VMAF analysis for Netflix 5G trace with ABR: Pensieve$^{+}$ and mode: VoD

### 4.1 Scenario A

**Main takeaway across traces:** VoD performance differences stem primarily from implementation-level design choices rather than CC algorithm selection. Identical CCs yield significantly different VMAF (up to 20-point variations) due to architectural factors including programming language efficiency, concurrency models, and pacing implementation. The trace characteristics determine which implementation weaknesses become pronounced—stable traces mask inefficiencies while volatile networks expose them.

We begin with the stable, high-throughput Netflix 5G trace, serving as a baseline to reveal core architectural efficiencies and overheads across implementations. For this trace (Figure 3), AIOQUIC shows similar VMAF scores with both Cubic and Reno. The main difference lies in variability: Cubic exhibits slightly higher standard deviation, reflecting its more aggressive probing under fluctuating bandwidth. Since AIOQUIC does not implement HyStart [3], Cubic's behavior leads to transient unfairness among concurrent flows, as also visible in the fairness results (Figure 9).

With MVFST, we observed comparable VMAF scores across various CCs, including BBR, BBR2, Cubic, Reno, Copa, and Copa2. Notably, MVFST consistently outperformed AIOQUIC in terms of VMAF scores when using the same CC (Cubic and Reno), with an average improvement of 6 units. This difference exceeds the Just Noticeable Difference (JND) threshold [54], implying that quality improvement with MVFST is perceptible to viewers. MVFST's superior performance lies in its C++ implementation, which offers significant advantages over AIOQUIC's Python-based design. C++ enables more efficient low-level resource management [8], including faster packet processing, reduced latency, and lower packet loss rates. These optimizations contribute to more stable network conditions, allowing MVFST to deliver higher and more consistent video quality, as reflected in the improved VMAF scores. This is evident in Figure 13 (left), where AIOQUIC demonstrates an average RTT that is 4× higher compared to MVFST, as well as other QUIC implementations.

With PICOQUIC, we observed comparable VMAF scores across most CCs, including BBR, Cubic, DCubic, Reno, and FastCC. However, BBR3 and Prague lagged slightly behind. BBR demonstrated better performance than BBR3, likely because BBR3's more aggressive bandwidth estimation mechanism can sometimes lead to faster throughput increases, but it may not always adapt smoothly to gradual bandwidth fluctuations. This can cause temporary instability in the network, which negatively impacts video quality. In contrast, BBR's more refined adaptation process allows for smoother adjustments, resulting in more stable throughput and higher VMAF scores. Prague's congestion avoidance strategy emphasizes pacing and stability, which may lead to under-utilization of bandwidth during gradual throughput increases, as it prioritizes avoiding bursts over aggressively ramping up the sending rate. With Cubic and Reno as CCs, PICOQUIC performed similarly to MVFST and exceeded the JND threshold compared to AIOQUIC. This improvement stems from the absence of Hystart in AIOQUIC and PICOQUIC's C++ implementation, which enables more efficient packet processing, lower latency, and more network stability than AIOQUIC's Python-based design.

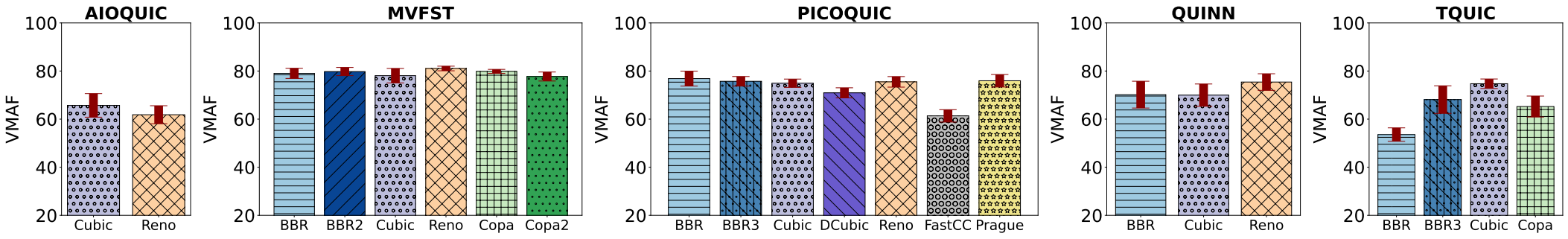

Fig. 4. Avg. VMAF analysis for Cascade trace with ABR: Pensieve$^+$ and mode: VoD

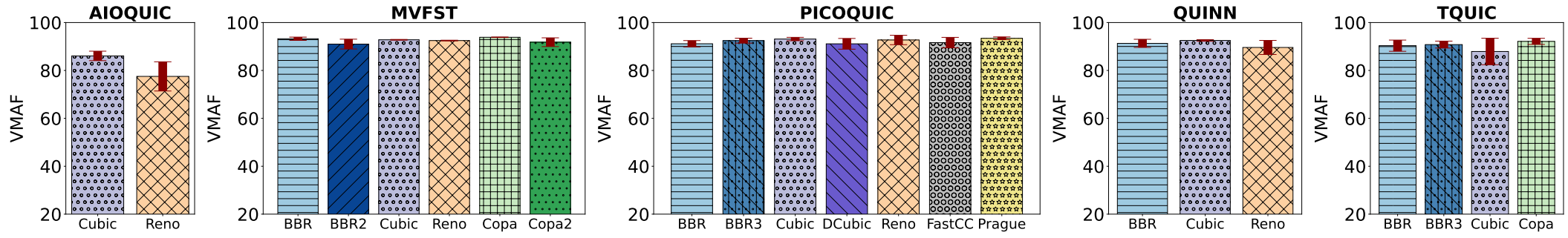

Fig. 5. Avg. VMAF analysis for LTE Belgium trace with ABR: Pensieve$^+$ and mode: VoD

With LSQUIC, we observed significantly lower VMAF scores for BBR compared to MVFST and PICOQUIC using the same CC. While LSQUIC is known for its faster RTT, it manages multiple connections using a single thread [51], which limits its ability to scale efficiently when handling multiple concurrent flows. With Cubic, LSQUIC shows a substantial improvement in VMAF scores, likely due to Cubic's aggressive ramp-up rate combined with LSQUIC's single-threaded design, which can be advantageous in high-capacity, low-latency networks like Netflix 5G trace. The single-threaded model introduces less overhead, allowing the connection to ramp up more efficiently under conditions where throughput steadily increases. While LSQUIC's performance still lags behind MVFST and PICOQUIC (which both utilize multi-threading for better scalability), it performs comparably with AIOQUIC, which is also multi-threaded. However, the fact that LSQUIC still performs relatively well, especially when compared to AIOQUIC, is surprising. This can be attributed to LSQUIC's advantage in lower RTT, as shown in Figure 13 (left) for the VoD mode, which implies faster response times. When using Adaptive CC, it dynamically switches between Cubic and BBR based on RTT. It uses Cubic when RTT is below 1.5 ms, contributing to higher VMAF scores than BBR but lower than Cubic. Despite its strengths in terms of RTT, LSQUIC's single-threaded model is a limiting factor when scaling to more complex use cases.

With QUINN, we observe comparable VMAF scores across different CCs, including BBR, Cubic, and Reno. In all cases, QUINN performs similarly to MVFST and PICOQUIC, surpassing the JND threshold when compared to AIOQUIC with the same CC. QUINN is implemented in Rust, and utilizes the asynchronous features of the `async-std` and `tokio` libraries, enabling efficient handling of concurrent connections in a non-blocking manner. This enables QUINN to outperform AIOQUIC, especially in scenarios leveraging Rust's memory safety and concurrency.

With TQUIC, we observe notably lower VMAF (3× below the JND threshold) when using BBR compared to QUINN, MVFST, and PICOQUIC with the same CC. This also holds for BBR3, where TQUIC underperforms relative to PICOQUIC. The primary reason is the lack of pacing support in TQUIC [66], which limits its ability to smoothly adjust sending rates and fully utilize available bandwidth. Without pacing, TQUIC sends packets in bursts, causing queue buildup, variable RTT, and fluctuating throughput, leading to lower perceptual quality.

A secondary factor is TQUIC's use of Hystart++ [16] with Cubic, which relies on ACK clocking—sending one ACK for every two packets—to control the sending rate. Operating without pacing, TQUIC depends heavily on ACK timing, producing bursty traffic and uneven bandwidth use under fluctuating conditions. Although Hystart++ improves responsiveness during slow start, it results in inconsistent throughput in multi-client streaming. Even with Copa, TQUIC remains below the JND threshold compared to MVFST, which leverages `folly::AsyncTransport` for asynchronous, smoother data handling.

With XQUIC, we observed significantly abnormal results compared to other QUIC implementations. Specifically, the VMAF scores using CCs such as Cubic, Reno, and Copa were notably worse, barely maintaining the lowest VMAF score for the given video. While some improvement was noted with BBR and BBR2, the improvements appear marginal when compared to other QUIC variants and these gains cannot be definitively attributed to their hybrid design or advanced CC mechanisms. One of the key issues observed with XQUIC is its poor scalability when handling multiple clients. Despite having a static parameter that limits the number of connection IDs to 2, XQUIC still accepts up to 5 connections, which could present a potential security vulnerability, as highlighted in [6]. The results presented in Figure 3 for XQUIC were obtained after adjusting the number of connection IDs to 5; however, no notable difference was observed, even in the absence of this modification. This is because XQUIC rewrites the global HTTP/3 callback context for each new connection [5], which means it only retains access to the most recently registered callbacks. This behavior significantly impacts the latency for subsequent requests, as only the latest request's callback is active, causing increased RTT for other simultaneous connections. This is particularly evident in the substantial RTT increases observed in XQUIC, as illustrated in Figure 13 (left). Additionally, XQUIC's handling of SSL handshakes is not optimized for concurrent requests. When multiple clients initiate requests simultaneously, XQUIC fails to properly schedule SSL handshakes [7], resulting in long initial RTTs as the server struggles to process multiple connections. This inefficiency is further exacerbated under high client concurrency, where the performance degrades noticeably.

LSQUIC and XQUIC showed inconsistent server performance due to poor multi-client handling. LSQUIC's single-threaded design and XQUIC's issues with HTTP/3 callback context rewriting and SSL handshake management caused significant degradation under concurrency, leading to low VMAF scores. Owing to these scalability limits and unfair comparability in multi-client scenarios, we excluded them from further analysis.

**Netflix 5G Trace Takeaway:** The high-throughput, stable nature of Netflix 5G (45 Mbps average) favors implementations with efficient resource management. MVFST and QUINN excel due to async architectures that maximize bandwidth utilization, while AIOQUIC's Python overhead and TQUIC's lack of pacing create significant performance gaps. This trace reveals how implementation efficiency becomes the differentiator in favorable network conditions.

Next, we examine the Cascade trace, which features mostly stable throughput with occasional drops, highlighting how each QUIC+CC adapts to abrupt rate changes and how pacing and delay sensitivity affect QoE. The results for the Cascade network trace with the same setup as before are presented in Figure 4. The outcomes with MVFST and AIOQUIC as the QUIC server implementations closely align with those observed for the Netflix 5G trace, showing similar VMAF trends across various CCs. With PICOQUIC as the QUIC server, the results also resemble those from the Netflix 5G trace but exhibit a few key differences. Specifically, BBR and BBR3 perform comparably for the Cascade trace. This similarity arises due to the stable throughput of the trace (see Figure 2), except for sudden drops. BBR and BBR3 can accommodate such drops effectively due to their sensitivity to delay and pacing mechanisms. Prague performs similarly to BBR for the same reasons, as its CC strategy is well-suited to handling steady throughput interspersed with occasional drops. Even Cubic and Reno achieve similar VMAF scores as BBR, BBR3, and Prague due to their conservative congestion window management during steady throughput. In contrast, DCubic lags slightly behind due to its less aggressive recovery behavior after throughput drops. FastCC lags behind considerably in terms of VMAF by 2× the JND threshold compared to other CCs. This gap arises from FastCC's poor adaptation to sudden throughput drops, as its aggressive rate probing and bandwidth-maximizing design suit only stable conditions.

With QUINN, we observed a noticeable drop in VMAF scores for all CC compared to the same CC implemented in MVFST and PICOQUIC. This drop in performance is consistent across the Cascade network trace and is 1× lower in terms of the JND threshold. The primary factor contributing to this difference lies in QUINN's use of Rust's `asyncio` module, which enables non-blocking operations for handling multiple concurrent connections. While non-blocking design offers advantages in terms of concurrency and efficiency under stable network conditions, it faces challenges in adapting to sudden throughput drops. This is because the asynchronous architecture may introduce delays in responding to rapid changes in network capacity, leading to brief periods of under-utilization or instability, which adversely impact video quality. Interestingly, with Reno, we observe a slight improvement in VMAF scores compared to BBR and Cubic when using QUINN. Reno's more conservative CC approach prioritizes stability and minimizes the likelihood of overshooting available bandwidth during throughput drops. This behavior aligns well with QUINN's non-blocking nature, as Reno's steadier data transmission rate allows the server to better handle asynchronous events without introducing additional delays or losses.

With TQUIC, contrary to the Netflix 5G trace, BBR3 performs substantially better than BBR, with Copa performing similarly to BBR3 in terms of VMAF. This improved performance of BBR3 in the Cascade trace can be attributed to its more aggressive bandwidth estimation mechanism, which helps it react more effectively to sudden throughput fluctuations, thereby increasing the VMAF of a video streaming session. Copa, similarly, balances throughput and delay, allowing it to achieve comparable VMAF scores to BBR3 by minimizing delay variance during network changes. Notably, in the Cascade trace, the performance of Cubic with TQUIC is on par with MVFST and PICOQUIC with the same CC. The Cascade trace presents a more stable network, where Cubic's aggressive ramp-up behavior, combined with TQUIC's incremental adjustments via Hystart++ and ACK clocking, is more effective. In this environment, TQUIC can maintain stable throughput and adjust incrementally to fluctuations without overwhelming the network or creating large queues, leading to more favorable VMAF scores.

**Cascade Trace Takeaway:** The stable baseline with sudden drops in Cascade exposes rate adaptation capabilities. Implementations with smooth pacing (MVFST, PICOQUIC) maintain quality during transitions, while aggressive CCs like FastCC and architectures with delayed response (QUINN's async model) struggle with abrupt changes. This trace highlights the importance of graceful degradation mechanisms.

Finally, we analyze the LTE Belgium trace, which features moderate average throughput and low standard deviation. This trace allows us to observe how implementations behave when the available bandwidth is sufficient but subject to smaller, frequent variations. The results for the LTE Belgium network trace, presented in Figure 5, show that, despite the trace's high fluctuations, the relatively low std (approximately 5 Mbps on average) makes the network conditions conducive for all CCs. Consequently, across all QUIC implementations, we do not observe major differences in terms of VMAF scores, except for AIOQUIC, where Reno provides slightly lower VMAF scores compared to Cubic. The performance of MVFST, PICOQUIC, QUINN, and TQUIC is similar, even though different CCs are used, primarily due to the easier nature of the LTE Belgium trace. An interesting observation from the LTE Belgium trace is that, despite the average throughput being around 20 Mbps per client, all clients are downloading segments encoded at the highest bitrate level of 40 Mbps. This is made possible by variable bitrate (VBR) encoding and a large playback buffer in VoD, where 40 Mbps represents the average bitrate across frames, allowing some segments to be encoded at lower bitrates based on content complexity [9]. The average segment size of 18 Mb for the 40 Mbps bitrate encoded video is sufficient to be delivered over the LTE Belgium trace with its 20 Mbps throughput. Notably, Pensieve$^{+}$ efficiently uses the segment size in its decision-making process, avoiding confusion with the average encoded bitrate. Consequently, the low std and moderate fluctuations in the LTE Belgium trace make it easier for the ABR to perform efficient bitrate decisions, allowing it to consistently request higher-quality segments, which is a key factor contributing to the improved VMAF scores observed with all QUIC implementations. Cascade and Netflix 5G trace are impacted by their low throughput in the first 40 seconds, leading to lower VMAF scores in comparison with LTE Belgium trace.

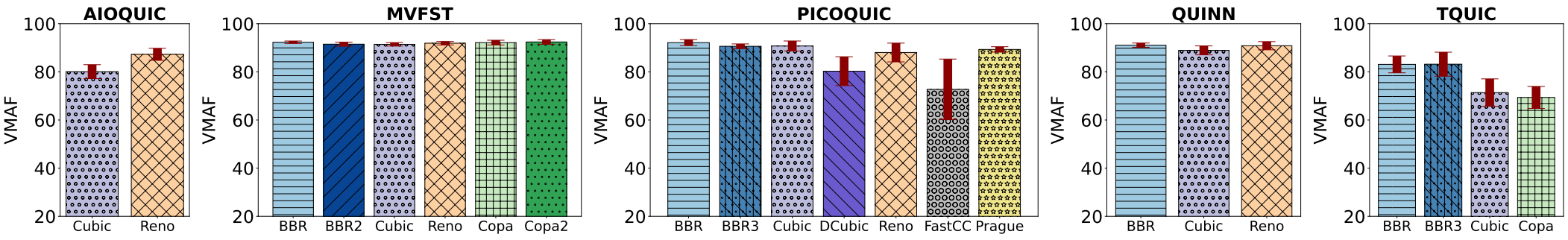

Fig. 6. Avg. VMAF analysis for Netflix 5G trace with ABR: Pensieve$^{+}$ and mode: LLL

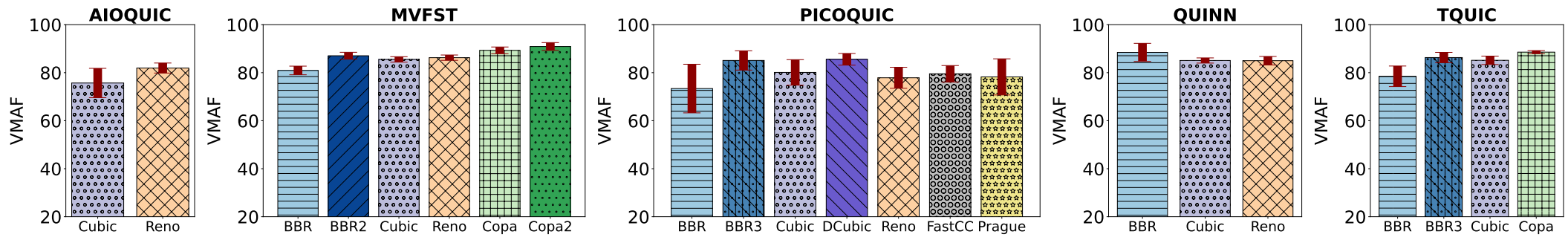

Fig. 7. Avg. VMAF analysis for Cascade trace with ABR: Pensieve$^{+}$ and mode: LLL

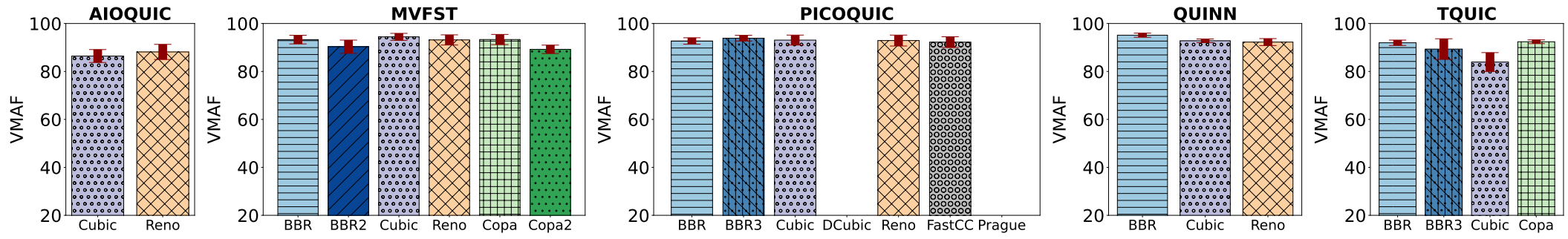

Fig. 8. Avg. VMAF analysis for LTE Belgium trace with ABR: Pensieve$^{+}$ and mode: LLL

**LTE Belgium Trace Takeaway:** Despite high fluctuation amplitude, the adequate average throughput (20 Mbps) and low standard deviation (5 Mbps) of LTE Belgium mask implementation differences, allowing even less efficient stacks to achieve high VMAF. The trace demonstrates that sufficient headroom can compensate for architectural inefficiencies, with VBR encoding and smart ABR decisions becoming the dominant factors.

## 4.2 Scenario D

In this subsection, we examine LLL streaming and start with the high-throughput Netflix 5G trace, followed by the more variable Cascade and LTE Belgium traces to test adaptation efficiency.

**Main takeaway across traces:** The constrained buffer (6s) and smaller segments (2s) in LLL streaming fundamentally alter CC behavior compared to VoD. Delay-sensitive CCs (BBR variants, Copa) generally excel, while aggressive ramp-up algorithms (Cubic) struggle with buffer constraints. Implementation architecture becomes even more critical as timely packet delivery directly impacts playback continuity.

The results for the Netflix 5G network trace are presented in Figure 6. With AIOQUIC, Reno outperforms Cubic in terms of VMAF exceeding the JND threshold. The low buffer size (6s) in LLL mode is not ideal for Cubic's aggressive ramp-up behavior. Cubic tends to overestimate available bandwidth [42] during its ramp-up phase, which leads to excessive sending rates and, ultimately, packet loss or delays before it can properly adjust to the available bandwidth, resulting in sub-optimal video quality. With MVFST, we observe similar VMAF scores no matter the CC being used. Notably, with Cubic and Reno, MVFST exceeds the JND threshold by ~2× when compared to AIOQUIC with the same CC. This observation mirrors the results presented in Section 4.1, where MVFST consistently outperformed AIOQUIC. The primary reason for this discrepancy can again be attributed to implementation reasons as discussed in Section 4.1. This is also demonstrated in Figure 13 (right), where AIOQUIC offers 3× higher RTT in comparison to MVFST as well as other QUIC implementations.

With PICOQUIC, the results mirror that of VoD mode in Section 4.1 with few key differences. For BBR, BBR3, Cubic, Reno and Prague, we observe similar VMAF scores and higher VMAF scores with Cubic and Reno in comparison to AIOQUIC with same CC due to similar reasons as highlighted before in VoD mode. The key difference with PICOQUIC is when DCubic and FastCC are utilized as CC, they lag behind in terms of VMAF by 1× and 2× the JND threshold, receptively, compared to Cubic. This performance gap can be attributed to the combination of smaller segment sizes and the limited buffer capacity in LLL mode, which intensifies the challenges of managing throughput fluctuations. DCubic's conservative ramping fails to fully utilize bandwidth, while FastCC's aggressive probing increases packet loss and delay. Consequently, both slow and erratic rate adjustments cause noticeable quality degradation.

With QUINN, consistent VMAF scores are observed across all CCs, closely resembling the results seen with MVFST and PICOQUIC, mirroring the results in VoD mode. QUINN's superior performance with Cubic and Reno, in comparison to AIOQUIC using the same CC, can again be attributed to its use of the efficient asynchronous library in Rust.

**Netflix 5G Trace Takeaway:** High-throughput stability favors delay-based CCs in LLL mode. Conservative algorithms (Reno) outperform aggressive ones (Cubic) in Python implementations, while efficient async architectures (MVFST, QUINN) maintain performance across CC variants.

With TQUIC, the results are the complete opposite of those observed in VoD mode. We suspect that the lack of pacing support in TQUIC, combined with the delay-based nature of BBR and BBR3, may align better with the limited buffer capacity in LLL mode. Our measurements show BBR and BBR3 achieve 30% higher VMAF compared to Cubic in TQUIC, which we attribute to their more conservative bandwidth probing that appears to cause less packet loss in buffer-constrained environments. On the other hand, Cubic and Copa perform similarly but lag behind in terms of VMAF by approximately 1× the JND threshold as compared to BBR and BBR3, likely due to their more aggressive nature that may be less suitable for LLL's constrained buffer conditions.

The results for the Cascade network trace are presented in Figure 7. With AIOQUIC, the results mirror that of Netflix 5G trace. When MVFST is used, we observe slightly lower VMAF with BBR, while Copa2 provides slightly higher VMAF in comparison to other CCs such as BBR2, Cubic, Reno, and Copa. The LLL mode requires quick and smooth adjustments to bandwidth, with minimal delays and packet loss. This is challenging for BBR, which relies on precise bandwidth estimation and pacing to manage throughput. BBR's aggressive bandwidth probing can lead to temporary over-estimations of available bandwidth, causing bursts of data that is detrimental to LLL mode. This results in slightly lower VMAF scores for BBR. In contrast, Copa2's more balanced approach to managing throughput and delay helps maintain a smoother video streaming experience, reducing the likelihood of sudden quality drops and contributing to the higher VMAF observed with Copa2.

With PICOQUIC, BBR again underperforms for the same reasons as MVFST+BBR. In contrast, BBR3 and DCubic achieve the highest VMAF scores. BBR3's switch to Hystart when RTT exceeds 250 ms improves adaptability to throughput fluctuations in LLL mode. Similarly, DCubic's delay sensitivity and Cubic-derived ramp-up optimize rate adjustment for low buffers and small segments.

Cubic, Reno, FastCC, and Prague perform comparably but slightly trail BBR3 and DCubic due to less effective bandwidth adaptation in LLL conditions.

With QUINN, surprisingly, we observe slightly higher VMAF scores when using BBR in comparison to MVFST and PICOQUIC with the same CC. However, on average, the VMAF scores are similar to those observed with PICOQUIC and MVFST when using Cubic and Reno. The reason for BBR performing slightly better with QUINN as compared to MVFST which also makes use of asynchronous connections is attributed to `asycnio` being lightweight in comparison to `folly::AsyncTransport`, which reduces overhead. As BBR relies heavily on precise bandwidth estimation and quick adjustments to maintain optimal throughput, the enhanced concurrency management in QUINN ensures better alignment with BBR's needs for LLL mode in stable throughput conditions, resulting in slightly improved VMAF scores.

With TQUIC, BBR again trails BBR3 in VMAF for reasons similar to PICOQUIC. BBR3, Cubic, and Copa achieve comparable VMAF scores, aligning with MVFST, PICOQUIC, and QUINN under the same CC. The Cascade trace's stable throughput favors Cubic's HyStart++ for smoother ramp-up without overestimating bandwidth, while Copa effectively balances throughput and latency, avoiding the aggressive probing issues seen in fluctuating traces like Netflix 5G.

**Cascade Trace Takeaway:** Sudden throughput drops expose CC adaptation capabilities. Hybrid approaches (BBR3, DCubic) excel, while pure delay-based (BBR) or loss-based (Cubic) algorithms struggle without implementation support. Lightweight async architectures (QUINN) provide the fastest adaptation to changing conditions.

The results for the LTE Belgium network trace are presented in Figure 8. Similar to VoD mode, the LTE Belgium trace yields consistent VMAF scores across most QUIC implementations, regardless of the CC. AIOQUIC continues to fall behind the JND threshold by 1× in terms of VMAF as compared to MVFST, PICOQUIC and QUINN with the same CC. Notably, exceptions arise with PICOQUIC where DCubic struggles to handle all five clients simultaneously, causing severe rebuffering for one client and resulting in a stalled video session. Similarly, Prague experiences significant rebuffering for two clients, leading to noticeable playback interruptions. These issues stem from the interaction between the CCs designs and the trace's unique characteristics. DCubic's less aggressive recovery mechanism, optimized for smoother bandwidth adaptation, under-performs in LTE Belgium trace, which has frequent but moderate bandwidth fluctuations, leading to persistent under-utilization of total capacity. This under-utilization becomes more pronounced in multi-client scenarios, where some QUIC flows severely underestimate the available bandwidth, which is further exacerbated by the limited playback buffer capacity in LLL mode, eventually leading to major rebuffering events. Prague, designed with delay-based sensitivity to minimize latency, overreacts to the trace's small but frequent throughput variations. This results in unnecessary rate reductions and inefficient packet pacing, compounding the problem in the context of small segment sizes and limited buffer capacity in LLL mode. These factors amplify the effects of consistently fluctuating throughput, causing playback interruptions. Unlike Netflix 5G and Cascade traces, which provide higher average throughput and more stable bandwidth, LTE Belgium exposes these weaknesses, particularly in multi-client setups. Notably, these issues are unlikely to occur in single-client scenarios [59], where resource contention and throughput estimation errors are minimal. With TQUIC, we observe a slight dip in VMAF, which can be attributed to the adaptation logic of Hystart++. While Hystart++ is designed to optimize the congestion window growth during the slow-start phase by avoiding premature timeouts, its reliance on RTT measurements to determine network congestion can cause suboptimal behavior in limited-buffer scenarios like LLL mode. Specifically, in traces with small but frequent throughput variations, such as LTE Belgium, Hystart++ might prematurely limit the sending rate to prevent latency spikes, underestimating the available throughput. This cautious behavior, while beneficial for reducing queuing delays, can lead to under-utilization of bandwidth, resulting in slightly lower VMAF scores due to occasional delays in delivering high-quality segments.

**LTE Belgium Trace Takeaway:** Adequate average throughput masks architectural inefficiencies but exposes CC algorithm weaknesses in multi-client scenarios. Conservative CCs (DCubic, Prague) fail catastrophically due to persistent bandwidth underestimation, while efficient implementations (MVFST, QUINN) maintain robustness across CC variants.

▹**LLL vs VoD Performance Insight:** Note that, on average across all scenarios, we observed higher VMAF scores in LLL mode compared to VoD mode. The key influencing factor here is the smaller segment sizes and durations (refer to Table 1) in LLL mode (exactly half that in VoD mode) combined with the same network trace. Additionally, the intelligent ABR decision-making of Pensieve$^{+}$ plays a key role by enabling the download of more segments in ultra-HD quality.

▹**Key Implementation Insights:**

- **Implementation Language Matters**: Low-level control in C++/Rust (MVFST, PICOQUIC, QUINN) consistently outperforms higher-level Python (AIOQUIC).
- **Concurrency and Architecture**: Async I/O architectures (QUINN, MVFST) outperform thread-based approaches (TQUIC), especially under multi-client load, by ensuring timely packet delivery.
- **Pacing is Fundamental**: Implementations without native pacing (e.g., TQUIC) exhibit bursty transmissions and poor bandwidth utilization in both VoD and LLL modes.
- **CC–Implementation Synergy**: The same CC algorithm behaves differently across implementations due to architectural choices (e.g., BBR excels in QUINN's lightweight async design but struggles in MVFST's batch processing).

### 4.3 QUIC Friendliness

This section examines server-side friendliness among multiple QUIC flows when serving 5 clients concurrently in both VoD and LLL modes for scenario **A**. "Friendliness" refers to fairness in achieved throughput and performance among concurrent flows. Results are consistent across network traces; we focus on the Netflix 5G trace for clarity. Due to similarities between VoD and LLL in flow fairness, LLL results (scenario **D**) are provided in Appendix C. The stable conditions of Netflix 5G help isolate performance differences among QUIC implementations and CC algorithms. LSQUIC and XQUIC results highlight unfairness from limited multi-client support. Notably, due to the increased sending rates offered by MVFST, QUINN, and PICOQUIC, the y-axis scales for these implementations have been adjusted accordingly. Clients are labeled C1–C5 for clarity.

Figure 9 presents the results for QUIC friendliness for all QUIC+CC combinations for Netflix 5G trace with Pensieve$^{+}$ deployed as the ABR on the client side. For AIOQUIC, we observe that Cubic is unfair to clients C2 and C3, due to Cubic's inherently aggressive CC behavior, which prioritizes rapid throughput recovery and a faster ramp-up during network fluctuations. This behavior results in transient unfairness, as Cubic may allocate more bandwidth to certain flows, causing clients C2 and C3 to receive lower shares of bandwidth compared to others during congestion periods. In contrast, AIOQUIC+Reno provides fair bandwidth allocation to all clients, resulting in similar average throughput across all clients. Reno's more conservative CC approach ensures a balanced distribution of available bandwidth, avoiding the transient unfairness observed with Cubic.

With TQUIC+BBR3 and TQUIC+BBR, slight differences in sending rates are observed across all clients. This can be attributed to their sensitivity to delay, which leads to short-term disruptions and adjustments in throughput, resulting in temporary unfairness towards some clients. We observe similar average throughput for Cubic and Copa, with Copa achieving higher throughput on average. However, Copa's higher throughput is accompanied by a larger std in bandwidth, indicating more variability in its performance across clients. Copa offers higher sending rates due to its aggressive approach to managing total capacity, which allows it to rapidly increase throughput compared to other CC like Cubic, especially in high throughput traces like Netflix 5G.

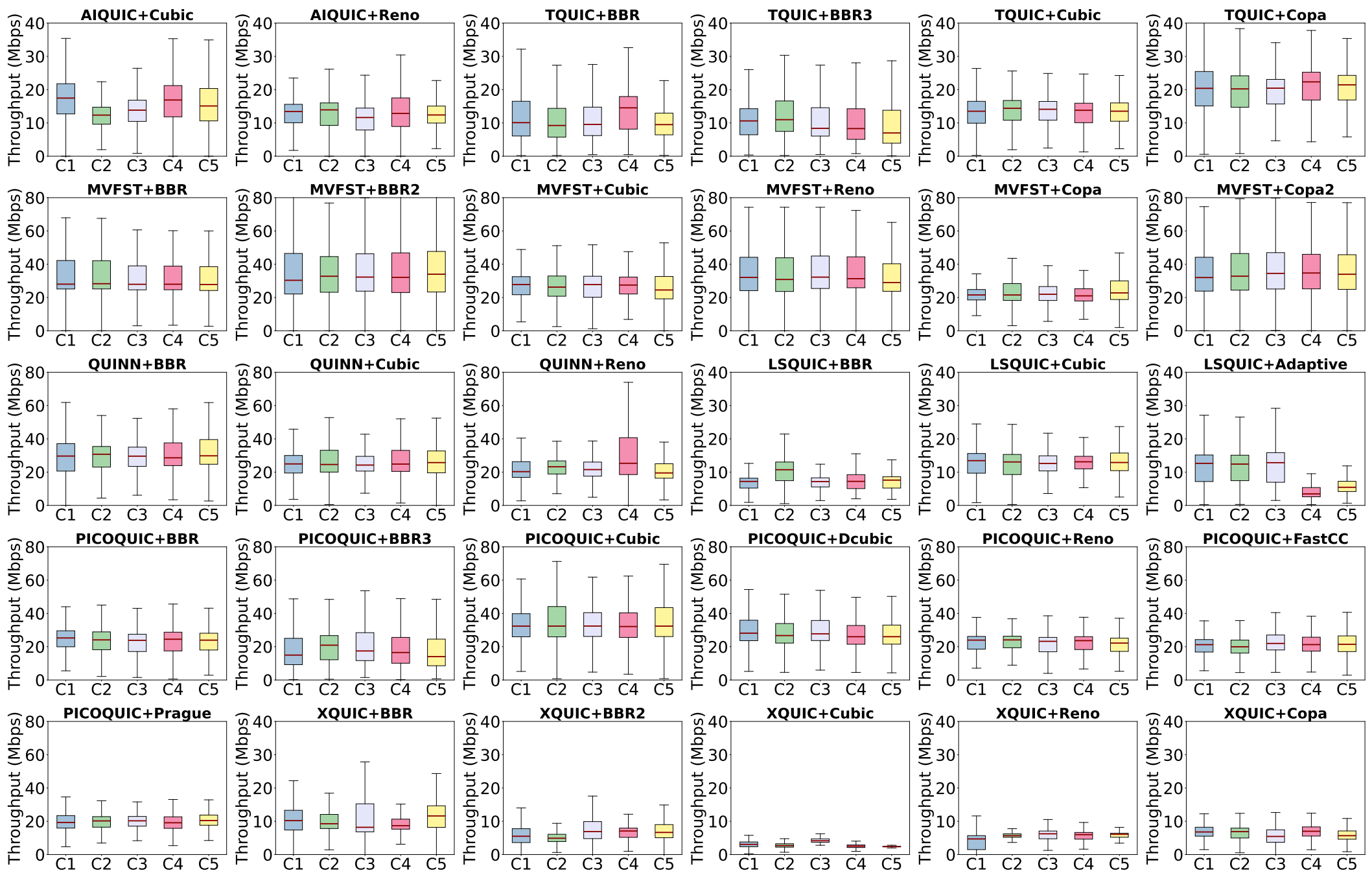

Fig. 9. QUIC friendliness for Netflix 5G trace with ABR: Pensieve+ and mode: VoD

Both MVFST and QUINN demonstrate consistent throughput across clients, regardless of the CC used, owing to their efficient asynchronous concurrency management. This design allows for optimal handling of multiple client connections, ensuring fair resource distribution and accurate bandwidth and delay estimations. With LSQUIC, we clearly observe extremely low throughput with BBR, though the average throughput is similar for all clients. With Cubic, the results are more balanced and noticeable unfairness arises when Adaptive is used. Specifically, clients C1, C2, and C3 experience significantly higher throughput, while C4 and C5 suffer from extremely low throughput. This disparity is due to LSQUIC handling all connections on a single thread, which leads to inefficient resource allocation and unfair bandwidth distribution among clients. Similar results are observed with XQUIC, where all clients experience extremely low throughput. This can be attributed to XQUIC's sequential SSL handshakes for each client and the re-writing of HTTP/3 context callbacks, which directly impacts the QoE for video streaming sessions, as highlighted in Section 4.1. PICOQUIC performs similar to MVFST and QUINN, by ensuring fairness among multiple flows for all clients, regardless of the CC algorithm used.

**QUIC Friendliness Takeaway:** Fairness in multi-client streaming is primarily determined by implementation architecture rather than CC algorithm selection. Asynchronous concurrency management (MVFST, QUINN, PICOQUIC) ensures consistent bandwidth distribution across clients regardless of CC choice, while single-threaded or sequential designs (LSQUIC, XQUIC) cause severe unfairness. CC algorithms introduce secondary effects—aggressive algorithms (Cubic) create transient unfairness, conservative ones (Reno) promote balance, and delay-sensitive variants (BBR) cause short-term fluctuations. The results underscore that scalable multi-client implementations must prioritize efficient connection handling over CC optimization to ensure fair QoE delivery.

### 4.4 Discussions

Our experiments show that both QUIC implementations and CC algorithms strongly impact adaptive video quality in VoD and LLL scenarios. MVFST, QUINN, and PICOQUIC outperform AIOQUIC due to lower RTTs and efficient asynchronous I/O. In VoD, Reno and Cubic improve quality via effective bandwidth management, while BBR's performance depends on network stability and pacing. In LLL, smaller segments combined with Pensieve$^{+}$ ABR reduce rebuffering and enhance experience. Fairness evaluations reveal that asynchronous implementations (MVFST, QUINN) maintain consistent throughput across clients, whereas single-threaded designs (LSQUIC, XQUIC) cause unfair distribution. CC affects fairness too: Reno balances sharing, Cubic can cause transient unfairness, and BBR introduces short-term variability. These results underscore the importance of transport optimizations, CC tuning, and client-aware concurrency for high-quality, fair, and scalable streaming.

## 5 Ablation Studies

Building on the limitations observed in our main analysis—where implementation architecture dominated performance, LLL mode revealed unexpected CC interactions, and multi-client fairness varied significantly—this section investigates critical corner cases that expose the boundaries of QUIC's capabilities. The stringent latency constraints of LLL mode amplify these limitations, making it an ideal testbed for probing: (1) how extreme 4G/5G volatility challenges even robust implementations, (2) whether router AQM strategies can mitigate architectural deficiencies, (3) if ECN marking improves fairness in unfair implementations, (4) whether throughput estimation errors compound implementation weaknesses, (5) how QUIC coexists with legacy TCP in mixed environments, and (6) which QUIC+CC pairings prove most resilient under stress. These studies provide crucial insights for deploying QUIC in real-world conditions where the ideal assumptions of our main experiments often break down.

### 5.1 Impact of High Volatility in 4G/5G

While 5G enables seamless 4K streaming, inconsistent performance—marked by highly fluctuating throughput, or *high volatility*—remains a challenge. Physical obstructions and network congestion disrupt connectivity, making it difficult for servers to maintain stable streaming. Figure 10 (left) presents the VMAF versus RD analysis for multiple QUIC implementations with their best-performing CC in LLL mode (refer to Section 5.6): AIOQUIC+RENO, MVFST+Copa, PICOQUIC+BBR3, QUINN+BBR, and TQUIC +BBR3, ensuring a fair comparison across QUIC implementations. The results presented are the average over 5 clients with Pensieve$^{+}$ used as the ABR on the client side. The average throughput and its std observed for this fluctuating 4G/5G trace are 137 Mbps and ±67 Mbps, respectively, with inter-variation time of 1s. This trace was made by combining several real-word 4G/5G traces from [58]. MVFST and QUINN demonstrate superior performance with the highest VMAF while providing a smooth streaming experience, with PICOQUIC trailing slightly and with some rebuffering. TQUIC falls short in VMAF with 3× the JND compared to MVFST and QUINN. Although AIOQUIC offers similar VMAF to PICOQUIC, it suffers from significant rebufferings. The efficient asynchronous non-blocking concurrency management in MVFST and QUINN enables them to adapt quickly to rapidly fluctuating network conditions. PICOQUIC and TQUIC with their multi-threaded design fall behind a bit. AIOQUIC, with its long RTTs combined with a slower response time because of its Python-based design, struggles with timely CC decisions, which leads to inefficient flow management and thus, it experiences high rebufferings in highly volatile network conditions.

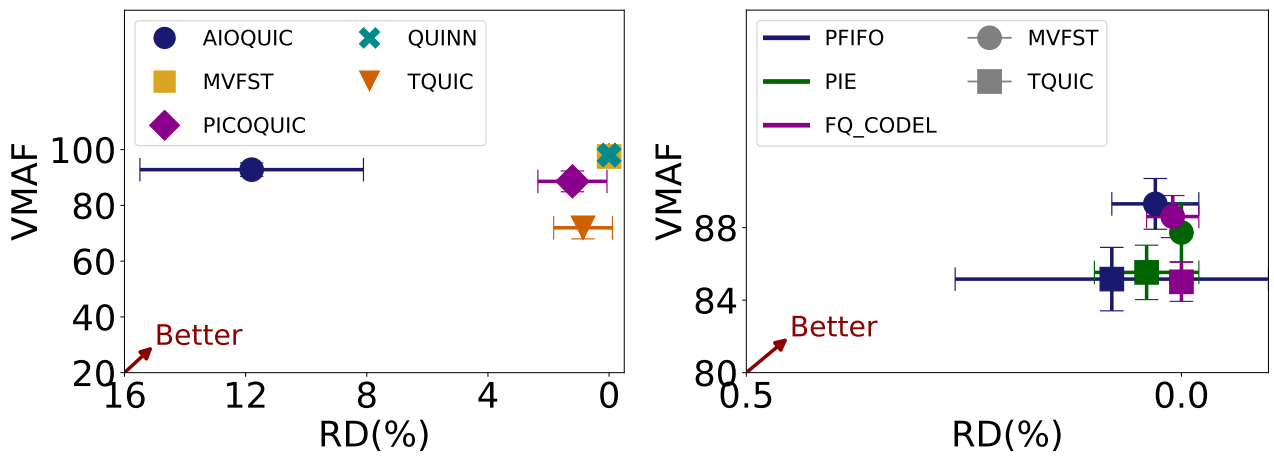

Fig. 10. Impact of High Volatility network trace (left) and AQM deployed at router (right)

## 5.2 Impact of router AQM strategy

Packet traversal often involves multiple hops, with routers using different AQM strategies. In this section, we examine the impact of the AQM strategy deployed at the router on QUIC performance. For better insights, we analyze the best and worst QUIC+CC combinations (MVFST+COPA and TQUIC+CUBIC) from Table 4 to answer: *i*) Does AQM affect QUIC performance? *ii*) If so, can it bridge the performance gap between QUIC implementations? The results presented focus on Pensieve$^{+}$ deployed as the ABR on the client side and the Cascade network trace, which presents challenges due to sudden throughput drops. The results for this setup are presented in Figure 10 (right), with an adjusted x-axis and y-axis for better clarity. Although the AQM strategy causes a minor 0.2% RD impact, it does not noticeably affect perceived video quality, indicating that QUIC CC performance is largely independent of the router's AQM.

## 5.3 Impact of ECN marking

Recent QUIC implementations have incorporated Explicit Congestion Notification (ECN) [34], allowing packets to carry congestion signals that enable CC algorithms to make more informed decisions. Among the implementations discussed in this paper, only LSQUIC, PICOQUIC, and QUINN fully support ECN [62]. However, due to the scalability issues of LSQUIC highlighted earlier, this section focuses on results from PICOQUIC and QUINN. These experiments deploy Pensieve$^{+}$ as the ABR algorithm on the client side, using the Cascade network trace in LLL mode. The Cascade trace, with its abrupt throughput drops, is particularly well-suited for analyzing the impact of ECN on performance. Moreover, we focus on the worst QUIC+CC combination in LLL mode: PICOQUIC+FASTCC and QUINN+Adaptive. As highlighted in Figure 11 (left), ECN provides slight improvements in terms of VMAF. Note that on average, the Cascade trace provides good throughput and this small impact will be amplified for extremely congested links.

## 5.4 QUIC Throughput Estimation

For the Netflix 5G and LTE Belgium network traces, we analyze the throughput estimated by QUIC-GO on client side, as these traces exhibit more dynamic behavior. Our goal is to assess whether inaccuracies in throughput estimation could contribute to the observed QoE differences, in addition to the influence of QUIC CC in each implementation. However, as shown in Figure 11 (right), the throughput estimates provided by QUIC-GO are consistent with the actual trace values, indicating that the observed performance differences are more likely due to factors other than client-side throughput estimation errors.

## 5.5 Interaction Between QUIC Flows and TCP Cross-Traffic

Although QUIC is becoming the preferred transport for media delivery, some providers still rely on TCP, making their coexistence important. We integrate five TCP Cubic flows into our SS-MC setup, each dedicated to a one-to-one streaming session to avoid TCP's head-of-line blocking. Results are shown for MVFST and QUINN in both VoD and LLL modes, using Cubic, the Netflix 5G trace, and Pensieve$^{+}$ ABR for consistency. As Figure 12 shows, QUIC traffic does not harm TCP throughput. The slightly higher TCP throughput reflects QUIC's multiplexing, which shares bandwidth across streams, but this trade-off is offset by reduced latency and improved handling of concurrent connections. Overall, QUIC and TCP coexist without significant degradation, even under high-demand scenarios.

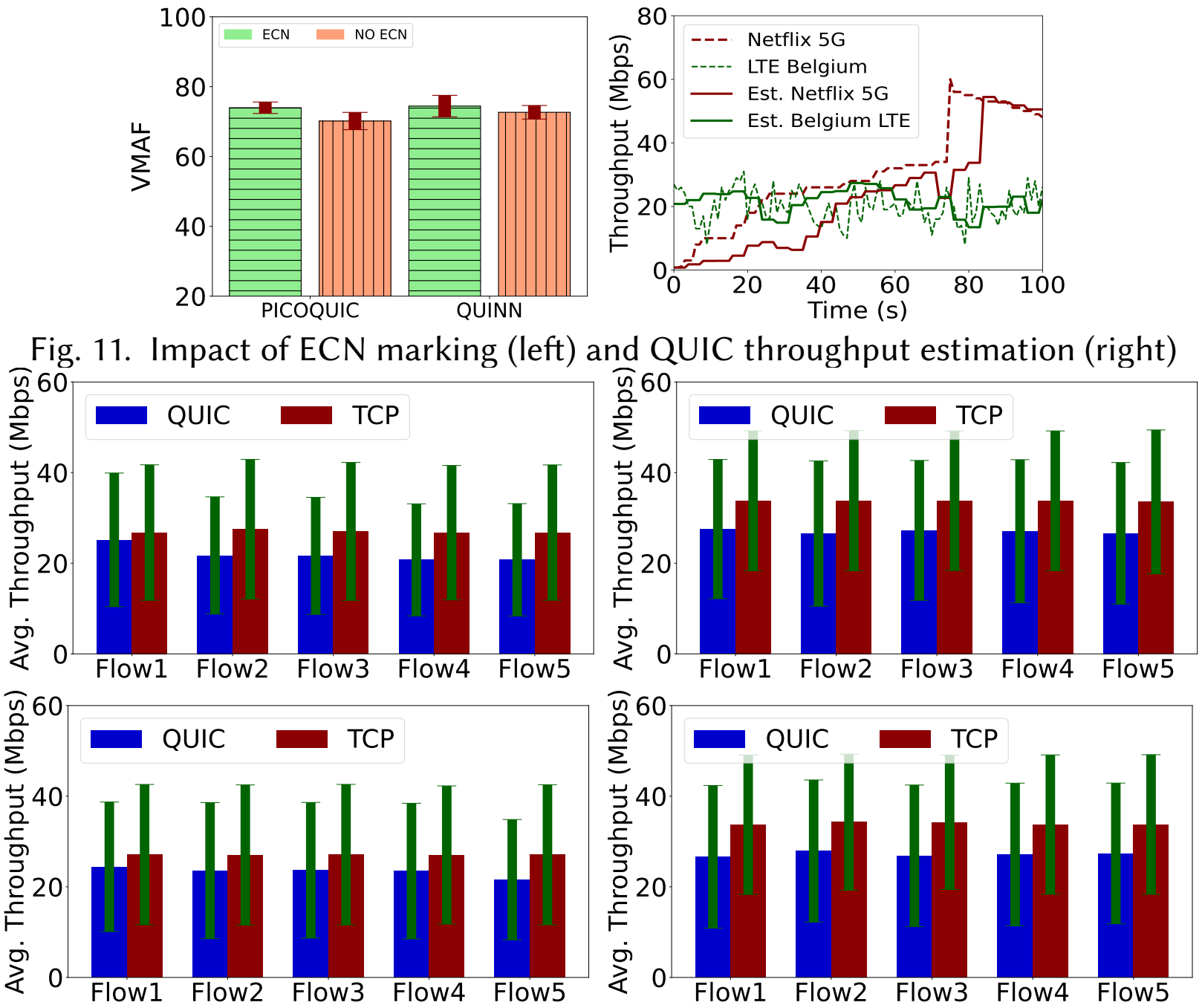

Fig. 11. Impact of ECN marking (left) and QUIC throughput estimation (right)

Fig. 12. QUIC vs TCP Flow Comparison for MVFST (left) and QUINN (right) for VoD (top) and LLL (bottom).

### 5.6 Best QUIC+CC

Table 4. QUIC VMAF summary.

| QUIC Implementation | CC | VoD | LLL |
|---|---|---|---|
| AIOQUIC | Cubic | 73.57 | 80.74 |
| | Reno | 68.57 | 85.79 |
| MVFST | BBR | **86.09** ↑ | 88.84 |
| | BBR2 | 84.74 | 89.62 |
| | Cubic | 83.96 | 90.47 |
| | Reno | 85.45 | 90.43 |
| | Copa | 85.25 | **91.59** ↑ |
| | Copa2 | 85.71 | 90.85 |
| PICOQUIC | BBR3 | 81.5 | 90.38 |
| | BBR | 82.74 | 85.57 |
| | Cubic | 84.08 | 87.98 |
| | DCubic | 81.39 | - |
| | Reno | 82.73 | 86.27 |
| | FastCC | 78.6 | 81.5 |
| | Prague | 81.63 | - |
| QUINN | BBR | 80.77 | 91.56 |
| | Cubic | 80.57 | 88.93 |
| | Reno | 80.16 | 89.37 |
| TQUIC | BBR | 66.72 | 84.52 |
| | BBR3 | 71.8 | 86.27 |
| | Cubic | 77.02 | 80.16 |
| | Copa | 75.31 | 83.44 |

Scenarios **A** and **D** (refer Table 1) serve as robust benchmarks for evaluating QUIC implementations and identifying the most effective CC for each implementation. Since modifying the CC in QUIC servers is straightforward, focusing on the best QUIC+CC combination ensures a level playing field, highlighting the true capabilities of each implementation without being constrained by suboptimal CC choices. Table 4 presents the average results of these comparisons over three distinct network traces. The selection criteria for the best CC prioritizes the highest VMAF, as we observed minimal or no RD with these scenarios, which allows VMAF to directly influence the QoE for video streaming. Enhancing the QoE aligns with the end-goal of any server implementation—delivering an optimal viewing experience. For VoD mode, the best QUIC+CC combinations are AIOQUIC+CUBIC, MVFST+BBR, PICOQUIC+CUBIC, QUINN+BBR, along with TQUIC+CUBIC, while for LLL mode, the optimal pairings are AIOQUIC+RENO, MVFST +COPA, PICOQUIC+BBR3, QUINN+BBR, and TQUIC+BBR3. For PICOQUIC, the FastCC and DCubic CC algorithms are excluded in LLL mode for the LTE Belgium trace due to unstable performance.

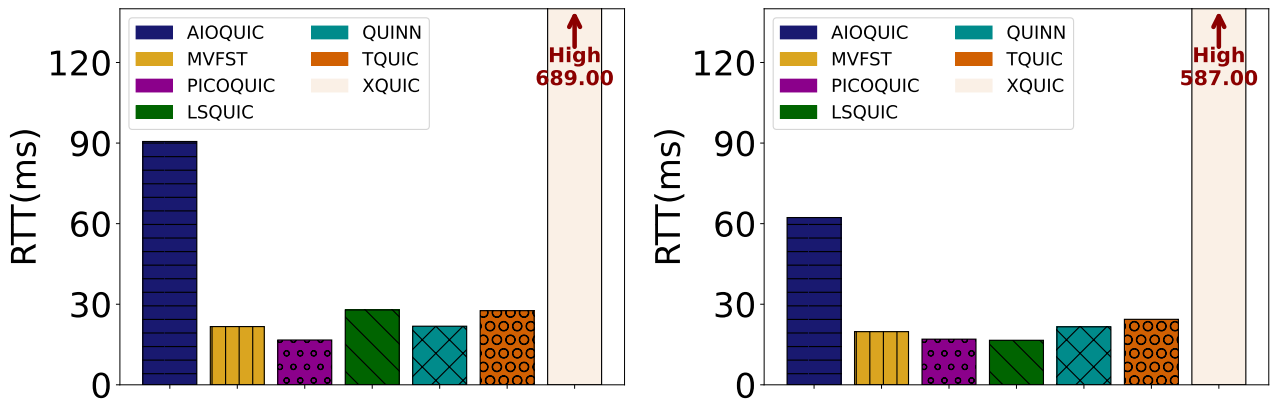

Fig. 13. RTT Comparison for QUIC implementations for VoD (left) and LLL (right).

MVFST+BBR achieves the best VMAF, with other MVFST-based CCs trailing slightly. PICOQUIC and QUINN show similar VMAF, with FastCC in PICOQUIC slightly lagging due to aggressive probing. For QUINN, all CCs perform similarly, while TQUIC and AIOQUIC generally underperform due to TQUIC's lack of pacing and AIOQUIC's long RTT, as shown in Figure 13. In LLL mode, TQUIC+Cubic lags due to issues with HyStart++.

Notably, the best performing QUIC+CC is MVFST+BBR for VoD mode and MVFST+Copa for LLL mode, as highlighted in green in Table 4. This underscores the robust design of MVFST and its efficient handling of concurrent client connections. This mode-specific analysis is necessary because VoD and LLL streaming have unique performance requirements—VoD emphasizes stable throughput for high video quality, while LLL prioritizes low latency. By isolating the best CC for each mode, the analysis ensures fairness and maximizes the insights gained from the comparative study of QUIC implementations.

## 6 Summary and Takeaways

Figure 13 shows the average RTT for each QUIC implementation, across all CCs, in both VoD and LLL modes. MVFST, LSQUIC, PICOQUIC, QUINN, and TQUIC exhibit similar RTT values. AIOQUIC has significantly higher RTT in both modes, with a slight reduction in LLL due to smaller segment sizes. XQUIC shows extremely high RTT, attributed to issues discussed in Section 4.1. LSQUIC, though low in RTT, uses a single thread for concurrent connections, making it unsuitable for scaled setups, similar to XQUIC.

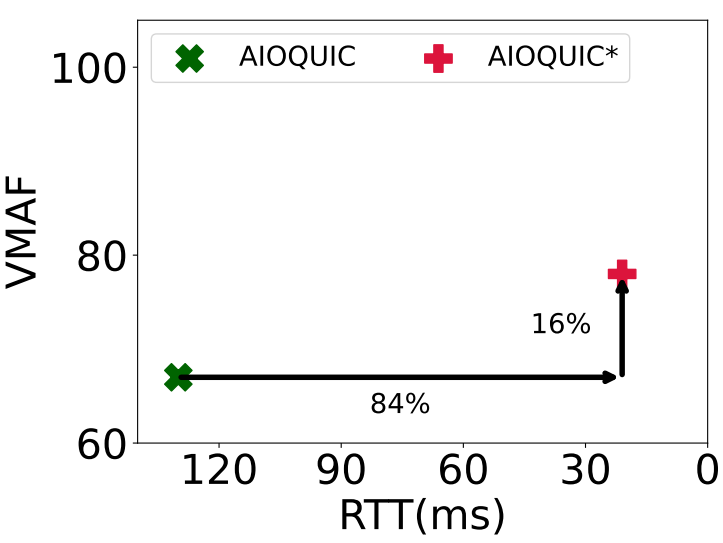

Fig. 14. Cross-Layer QUIC Server.

Our results emphasize the need for QUIC flow awareness and a cross-layer approach that adapts delivery rates based on content type and client requirements. Such strategies could improve performance by optimizing the interaction between CC and ABR mechanisms, enhancing bandwidth distribution and bridging gaps between QUIC implementations. These insights could be pivotal for advancing the media over QUIC (MoQ) [29] stack, enabling more efficient media delivery in diverse network environments. An additional experiment integrated the Connection Aware QUIC design [64], where the QUIC server knows the maximum video segment chunk size from the MPD or manifest file before establishing a connection. We modified the AIOQUIC+RENO configuration to set connection-wide and per-stream flow control limits to the maximum chunk size. This adjustment, referred to as AIOQUIC*, reduces latency and increases efficiency. Experimental results, shown in Figure 14 for Scenario **A** (see Table 1) using the Netflix 5G trace, demonstrate a 16% increase in VMAF and a 84% reduction in RTT compared to the unmodified version, significantly improving user QoE. This highlights the potential benefits of a cross-layer metric-sharing approach in improving QUIC performance. While this is an initial step, it requires further real-world evaluation but sets the foundation for future exploration in QUIC and MoQ advancements.

## 7 Limitations and Future Work

While we explored a range of scenarios in video streaming and media delivery, it is important to recognize the broad scope of challenges in this domain, and our study has several limitations. First, we focused on just seven QUIC implementations, despite approximately 15 available today. Some implementations, like QUICHE, are limited to one-to-one server-client setups and lack multi-client support, similar to LSQUIC and XQUIC. Features such as prioritization, 0-RTT, and connection migration were not addressed. QoE analysis was limited to 4K videos, as no official VMAF exists for 8K; future work will cover 8K, 360°, and volumetric videos with higher bandwidth and ultra-low latency demands. Client-side impacts of different QUIC implementations were not studied and will be examined in future work. We also plan to analyze MoQ under similar conditions, explore varying device resolutions and streaming modes, and extend our evaluation to real-world settings, including browser-specific effects on QUIC-based video streaming.

## 8 Conclusion

This study analyzes QUIC CC performance in video streaming, highlighting the impact of implementation differences, queuing strategies, content types, and client-side ABR schemes on viewer QoE. Our findings show that QUIC CC significantly influences ABR decisions, directly affecting viewer QoE. QUIC implementations in Rust/C++ generally offer faster response times, though server-side delivery can impact performance. Asynchronous connection handling, as seen in QUINN, is more efficient than multi-threading, like in TQUIC. Additionally, QUIC and TCP can coexist in the same network with good level of bandwidth fairness. The study underscores the need for tailored CC and ABR solutions for QUIC, as many current implementations are TCP-optimized. No single QUIC+CC implementation outperforms others universally, highlighting the need for future work on cross layer QUIC/ABR design and unified QUIC implementations such as MoQ, to overcome current limitations and enhance versatility across diverse scenarios. While this study focused on throughput-based performance, we acknowledge that network latency, loss characteristics, and concurrent traffic effects are important factors, which we plan to investigate in future work.

# Online Appendix

## A Ethics

In accordance with the guidelines set by the Human Ethics Research Committee (HERC) of our university, our research adheres to all ethical standards. All implementations utilized in this study are open-source and publicly available, ensuring transparency and avoiding any ethical concerns. Furthermore, our analysis was conducted without any attempt to link video streaming sessions to personal identities, maintaining strict compliance with ethical research practices.

## B Additional Scenarios

### B.1 Scenario B

Figure 15 illustrates the VMAF versus RD results for various QUIC implementations under the Netflix 5G trace for VoD mode. In this setup, 5 clients (C1–C5) use different ABR algorithms: C1:Pensieve$^+$, C2:Merina$^+$, C3:BBA2-C, C4:EXP, and C5:CON. The Netflix 5G trace was chosen because of its higher average throughput and lack of abrupt drops, creating a more ABR-friendly environment. For each QUIC implementation, the best QUIC+CC was chosen as highlighted in Section 5.6. As the Netflix 5G trace offers high throughput and the VoD mode features a large buffer capacity (60s), each ABR algorithm consistently manages to download ultra-HD segments. Consequently, across all QUIC implementations, there is no significant difference in VMAF among the selected ABRs. The only notable discrepancy is the slight rebuffering observed for C1 and C2 when using TQUIC with Pensieve$^+$ and Merina$^+$. These learning-based ABRs tend to request 4K segments, and if the server fails to respond promptly, stalls can occur. Overall, MVFST and QUINN deliver the highest VMAF, with their asynchronous design providing a performance edge by enabling efficient data handling and reducing delays. PICOQUIC and TQUIC follow closely behind, performing well but slightly less efficiently in comparison. AIOQUIC lags behind the others, likely due to its higher RTT, which can be attributed to less efficient concurrency handling in its Python-based implementation.

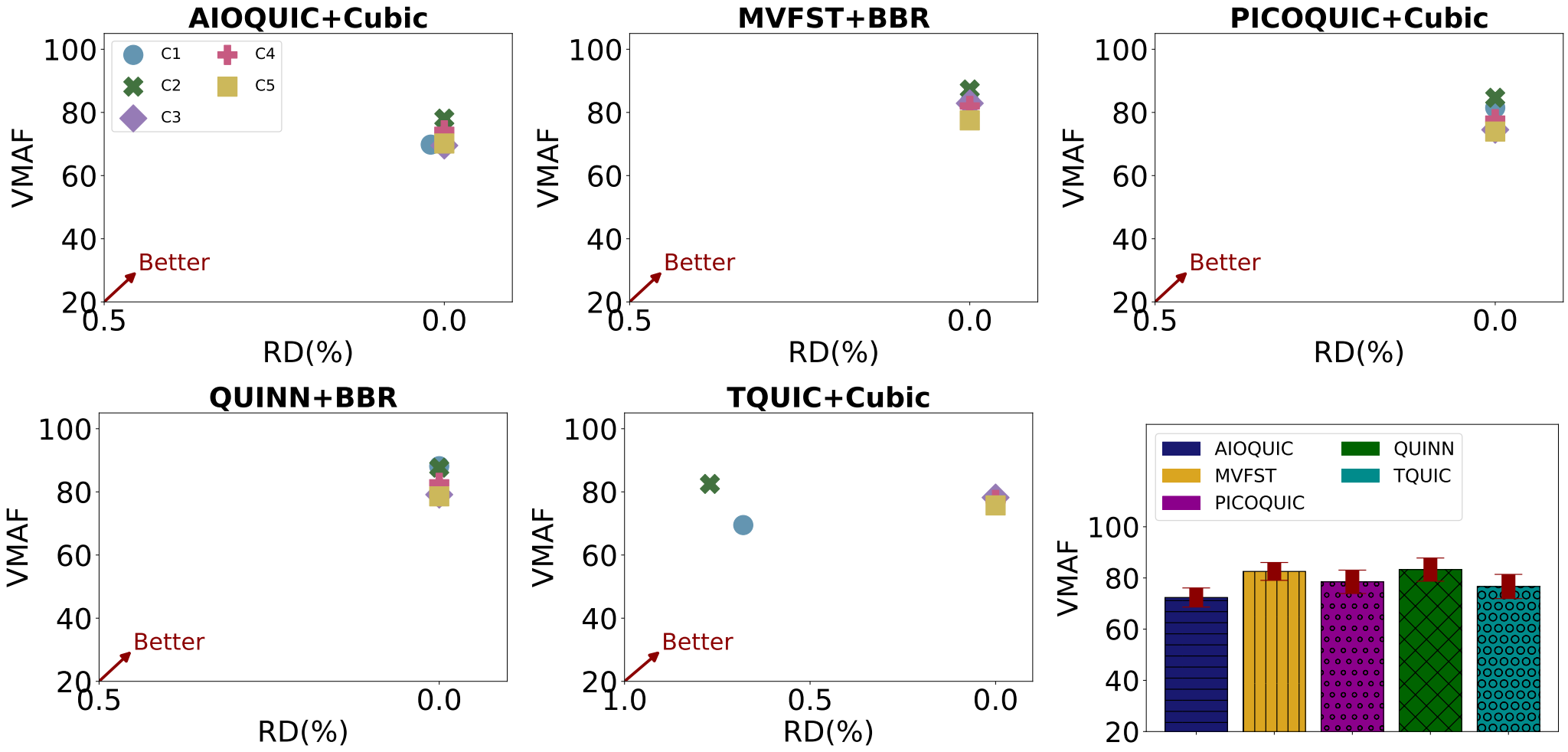

Fig. 15. QoE analysis for Netflix 5G trace with different ABR and mode: VoD

## B.2 Scenario C

Figure 16 illustrates the VMAF versus RD results for various QUIC implementations with Pensieve+ is used as the ABR on client side for VoD mode. In this setup, five clients (C1–C5) operate under distinct network conditions: C1:Netflix 5G, C2:LTE Belgium, C3:Cascade, C4:a constant 60 Mbps, and C5: a constant 5 Mbps, respectively, whereas the QUIC server is operating with constant 200 Mbps bandwidth. The results for MVFST, QUINN, and TQUIC are closely aligned, with C4 achieving the highest VMAF score, as expected, followed by C2, C1, and C3. C5 records the lowest VMAF score, which is unsurprising given its limited throughput. Interestingly, LTE Belgium yields a higher VMAF score than Netflix 5G, which may seem counterintuitive but can be attributed to the factors discussed in Section 4.1. AIOQUIC follows a similar pattern but consistently achieves lower VMAF scores. This can be attributed to AIOQUIC's Python-based implementation, which generally results in slower performance compared to the C++ or Rust implementations used by other QUIC versions. This is also evident with longer RTT times for AIOQUIC as seen in Figure 13. In terms of VMAF, PICOQUIC shows a similar trend to MVFST, QUINN, and TQUIC, but it is the only implementation to experience a small but noticeable rebuffering event for C3. This issue arises when the server initially sends more data, and then the sudden drop in throughput due to the Cascade trace significantly impacts Cubic's ability to adapt. This rebuffering event is absent in TQUIC, likely due to its lack of pacing support, which, in this scenario, actually helps mitigate the negative effects of fluctuating throughput.

## B.3 Scenario E

Figure 17 illustrates the VMAF versus RD results for various QUIC implementations under the Netflix 5G trace for LLL mode. We use the same setup discussed in Appendix B.1. The results for the LLL mode closely resemble those observed in the VoD mode, with a significant distinction. BBA2-C, as a purely buffer-based algorithm, is heavily constrained by the limited buffer capacity in the LLL mode. This limitation forces it to consistently download only low-quality segments, despite the fact that the average throughput of Netflix 5G is sufficient to support high-quality segment downloads, as evidenced by the VMAF scores achieved with other ABRs. As a result, the server finds it easier to accommodate high-quality segment requests from other clients since BBA2-C primarily requests minimal segment sizes. Consequently, no significant differences are observed across the QUIC implementations.

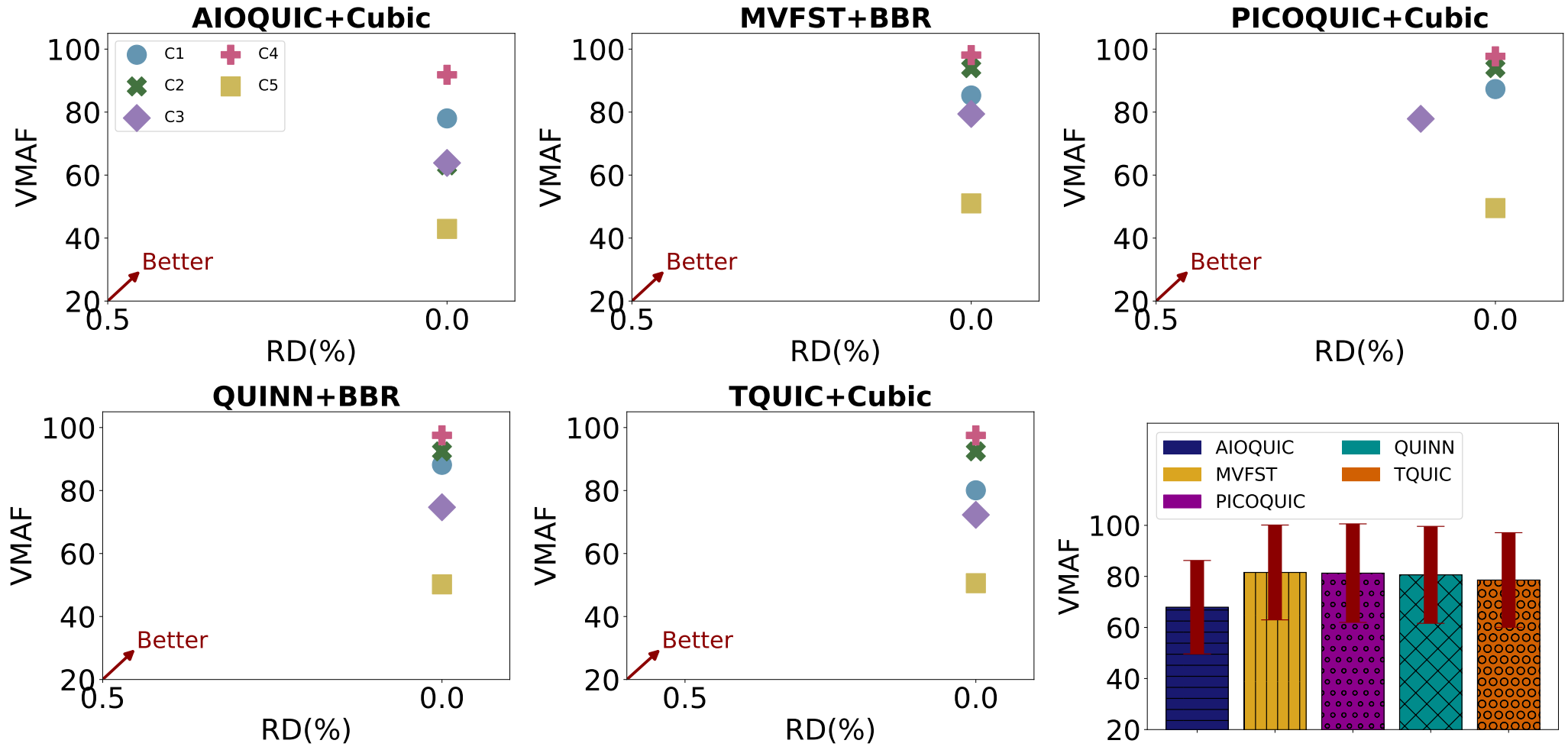

Fig. 16. QoE analysis for different network traces with ABR:Pensieve+ and mode: VoD

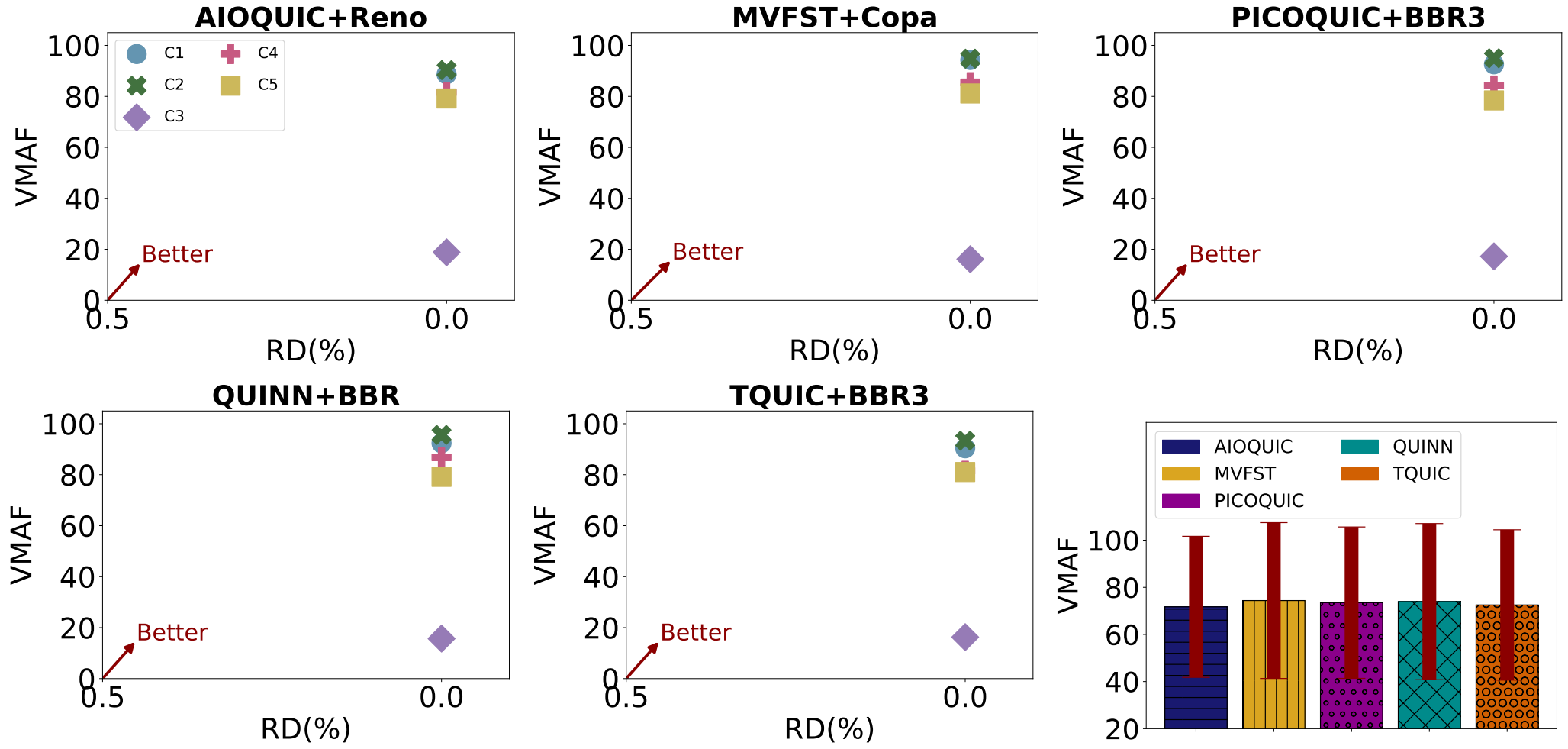

Fig. 17. QoE analysis for Netflix 5G trace with different ABR and mode: LLL

### B.4 Scenario F

Figure 18 illustrates the VMAF versus RD results for various QUIC implementations with Pensieve$^{+}$ is used as the ABR on client side for LLL mode. We use the same setup discussed in Appendix B.2. The results resemble that of VoD mode with higher VMAF as smaller segment sizes enable quicker adaptation to network conditions, improving the overall quality in LLL mode while still being constrained by the available throughput, similar to the discussion in Section B.1.

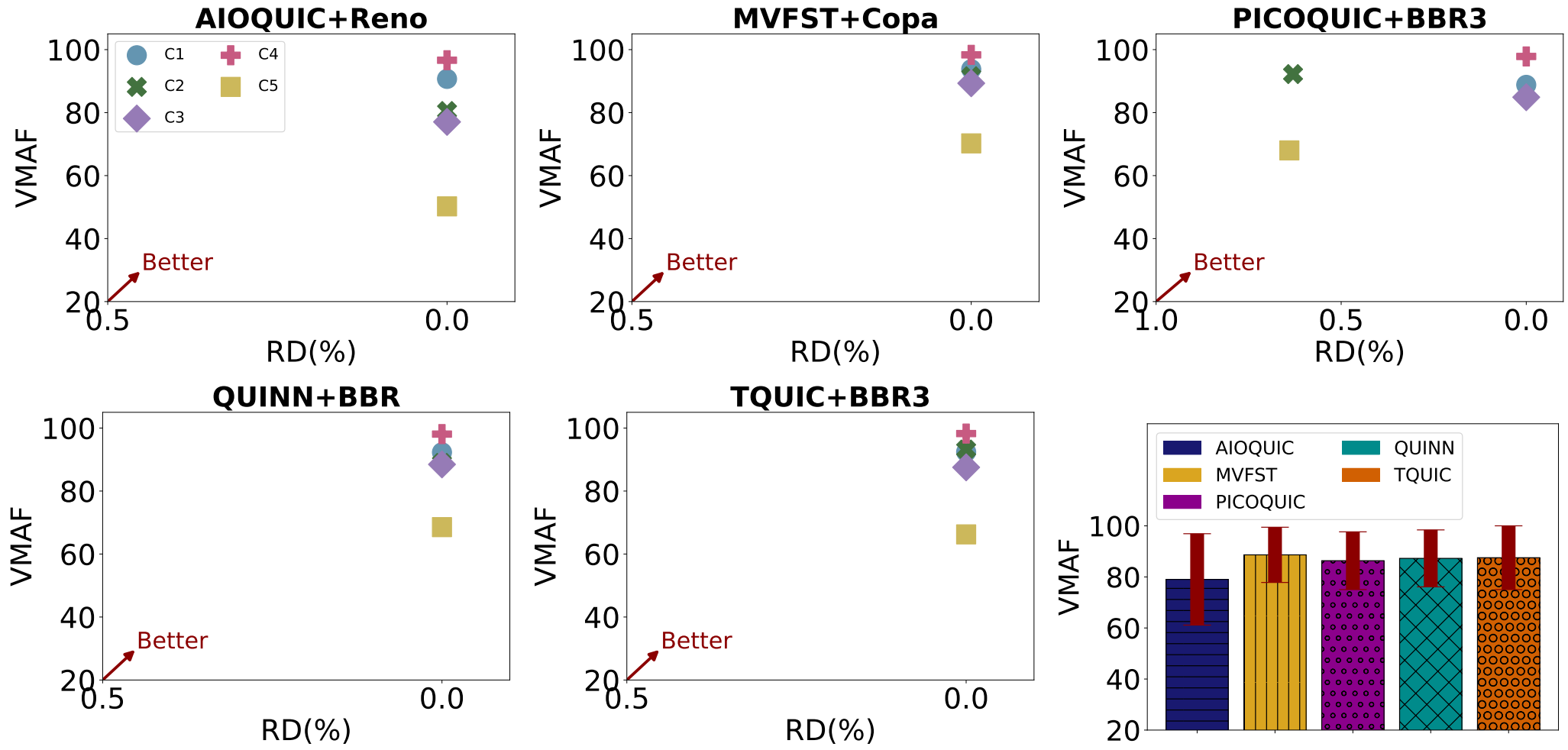

Fig. 18. QoE analysis for different network traces with ABR:Pensieve$^{+}$ and mode: LLL

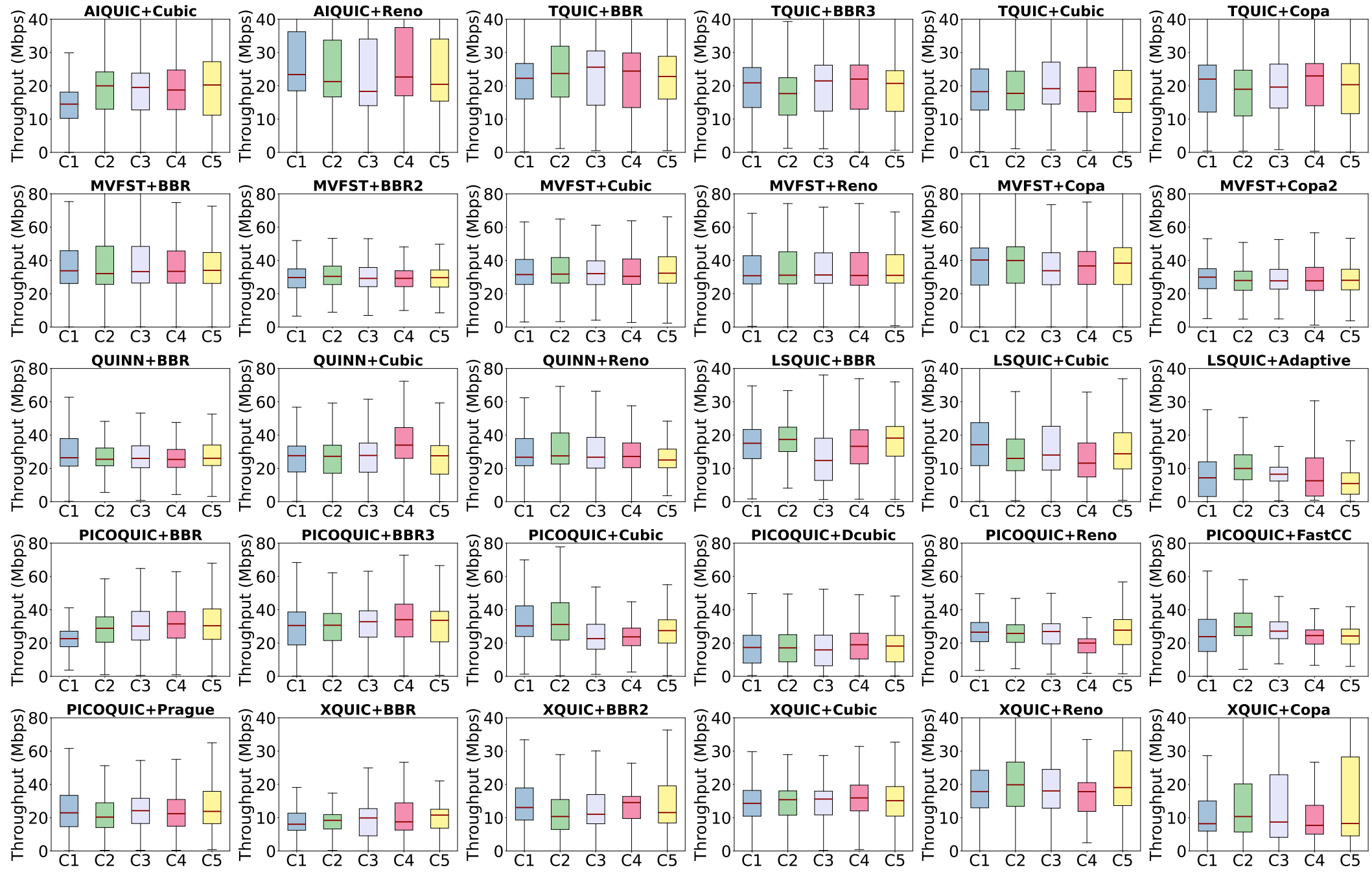

Fig. 19. QUIC friendliness for Netflix 5G trace with ABR: Pensieve⁺ and mode: LLL

## C QUIC Friendliness

Figure 19 illustrates the QUIC friendliness results for the Netflix 5G trace with Pensieve⁺ deployed as the ABR on the client side for Scenario D (refer to Table 1). The results align closely with those observed in VoD mode for Scenario A, underscoring recurring patterns in how QUIC implementations handle dynamic conditions. AIOQUIC+Cubic exhibits unfriendly behavior toward C1 due to its aggressive ramp-up, leading to transient unfairness among flows. However, AIOQUIC+Reno provides a more balanced performance, achieving the same average throughput as Cubic but with reduced competition and smoother flow interactions. All CC algorithms in TQUIC demonstrate relatively similar behavior, indicating that its design is inherently fair across flows. MVFST and QUINN emerge as robust implementations, maintaining fairness across all flows regardless of the CC algorithm used. Notably, MVFST+BBR2 and QUINN+BBR offer slightly lower mean throughput due to their conservative nature. PICOQUIC is also largely fair, though it struggles with Reno and FastCC, which exhibit less predictable behaviors under high variability. In the LLL mode, the frequent segment requests caused by low client buffer capacities and small segment sizes place additional stress on CC algorithms. This can amplify inefficiencies in overly aggressive or overly conservative algorithms, particularly during abrupt throughput fluctuations. XQUIC and LSQUIC continue to show instability and unreliability, primarily due to their lack of multi-client support. Their inclusion here highlights the significant performance gaps in implementations that lack scalability, further emphasizing the need for robust multi-client designs.

## D Experimental Setup Details.

Figure 20 provides a detailed overview of the extended Vegvisir setup for the Single Server-Multi Client (SS-MC) configuration. The server is equipped with a single network interface, *eth0*, which is responsible for transmitting data to the traffic shaper. The data flow is facilitated through a Docker bridge, acting as a gateway defined by Docker Compose networks. This bridge enables seamless communication between the server and the shaper, ensuring isolation and manageability. The shaper, central to traffic management, implements **Token Bucket Filters (TBF)** for regulating and shaping the flow of traffic. This ensures precise control over the rate at which data is transmitted, simulating network conditions such as delay, loss, and varying bandwidth. The shaper uses Priority First In First Out (PFIFO) as the default queuing discipline (qdisc).

The shaper directs traffic to five of its interfaces: *eth1*, *eth2*, *eth3*, *eth4*, and *eth5*, with each interface dedicated to a specific client. On the client side, another Docker bridge is employed to establish connectivity between the shaper and the clients. Each client runs in an isolated Docker container and is assigned its own dedicated network interface, *eth0*, to receive traffic from the shaper. This design ensures a controlled and reproducible environment for testing, with each client functioning as an independent entity. To enable seamless routing across the setup, IP routes are pre-configured using the `193.167.0.0/16` subnet. This routing table configuration ensures that all server, shaper, and client containers are interconnected within the same virtual network, allowing data to traverse efficiently through the pipeline.

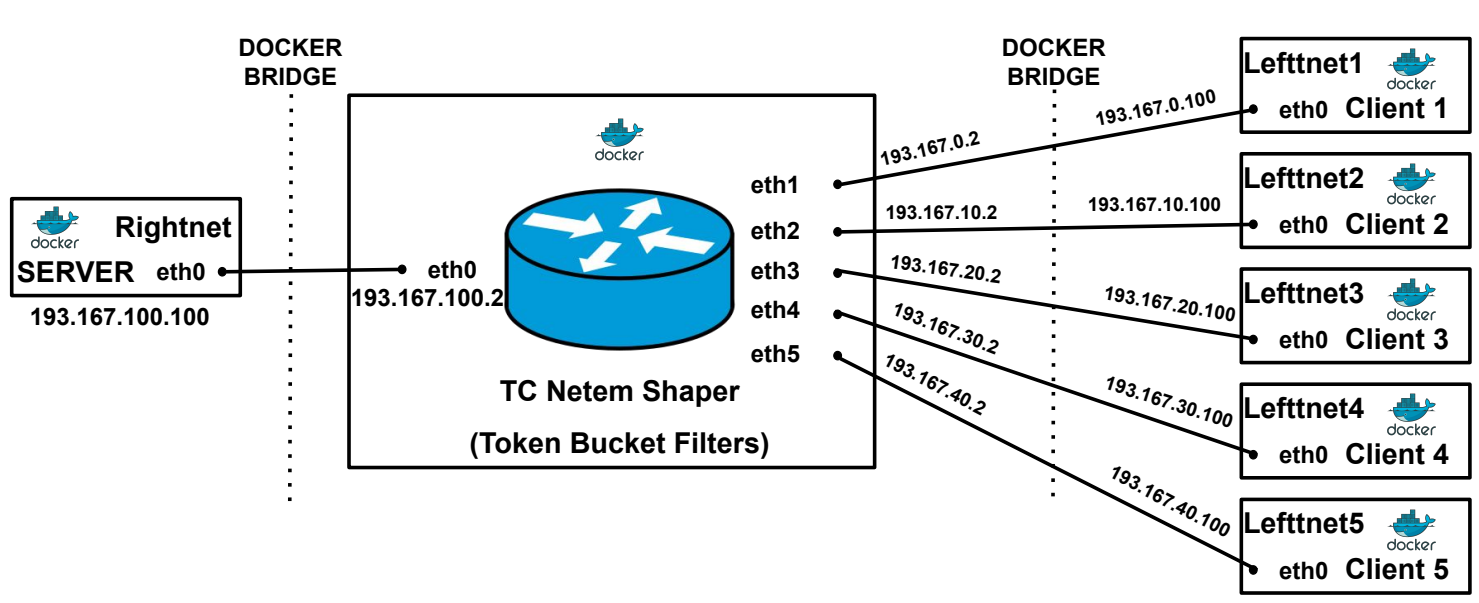

Fig. 20. Detailed Vegvisir Setup.

Table 5 provides a comprehensive overview of the buffer and size limits configured for each network interface within the traffic shaper. These configurations are tailored to manage the data flow effectively in the SS-MC setup. The shaper's *eth0* interface, which handles the flow of data from one server to all five clients, is assigned increased buffer and size limits to accommodate the higher traffic volume passing through it. This ensures minimal bottlenecks and smoother delivery to downstream client interfaces. Notably, delay and packet loss settings were not applied in this setup to preserve the natural behavior of QUIC. However, the burst size for the TBF is calculated based on the current transmission rate multiplied by the default latency value of 10ms. This dynamic calculation ensures appropriate queue sizes based on the transmission conditions. For experiments requiring ECN marking, a minimal 1% packet loss was introduced, as ECN inherently requires a loss event to trigger its marking mechanism. When testing other AQM strategies at the shaper like Proportional Integral Controller Enhanced (PIE) and Flow Queue CoDel (FQ_CoDel), the default configurations for these strategies were employed to maintain consistency and comparability across experiments.

Table 5. Shaper Details.

<table>
<tr><th>Interface</th><th>Buffer</th><th>Limit</th></tr>
<tr><td>eth0</td><td>300k</td><td>300k</td></tr>
<tr><td>eth1</td><td rowspan="5">100k</td><td rowspan="5">100k</td></tr>
<tr><td>eth2</td></tr>
<tr><td>eth3</td></tr>
<tr><td>eth4</td></tr>
<tr><td>eth5</td></tr>
</table>